\documentclass[usenatbib]{mnras}

\usepackage{graphicx}	
\usepackage{amsmath}	
\usepackage{amssymb}	

\newcommand {\kms} {\,{\rm km\,s}^{-1}}
\newcommand {\mo}{{\rm M}_\odot}

\title[The origin of SECCO~1]{Alone on a wide wide sea. The origin of SECCO~1, an isolated star-forming gas cloud in the Virgo cluster\thanks{Based on data obtained with the European Southern Observatory Very Large Telescope, Paranal, Chile, under the Programme 295.B-5013.}\thanks{Based on observations made with the NASA/ESA Hubble Space Telescope, obtained at the Space Telescope Science Institute, which is operated by the Association of Universities for Research in Astronomy, Inc., under NASA contract NAS 5-26555. These observations are associated with program GTO-13735.}\thanks{Based on observations made with the GTC, installed in the Spanish Observatorio del Roque de los Muchachos of the Instituto de Astrofísica de Canarias, in the island of La Palma.}}

\author[M.Bellazzini]{
M. Bellazzini$^{1}$\thanks{E-mail: michele.bellazzini@oabo.inaf.it},
L. Armillotta$^{2}$
S. Perina$^{3}$,
L. Magrini$^{4}$,
G. Cresci$^{4}$,
G.Beccari$^{5}$,
\newauthor
G. Battaglia$^{6,14}$, 
F. Fraternali$^{7,15}$,
P.T. de Zeeuw$^{8,9}$,
N.F. Martin$^{10,11}$, 
F. Calura$^{1}$,
\newauthor
R. Ibata$^{10}$,
L. Coccato$^{5}$,
V. Testa$^{12}$,  
M. Correnti$^{13}$
\\
$^1$ INAF - Osservatorio di Astrofisica e Scienza dello Spazio di Bologna, Via Gobetti 93/3, 40129 Bologna, Italy\\
$^2$Research School of Astronomy and Astrophysics - The Australian National University, Canberra, ACT, 2611, Australia\\
$^{3}$ INAF - Osservatorio Astronomico di Torino, Via Osservatorio 30, 10025 Pino Torinese, Italy\\
$^4$INAF - Osservatorio Astrofisico di Arcetri, Largo E. Fermi 5, 50125 Firenze, Italy\\
$^5$European Southern Observatory, Karl-Schwarzschild-Strasse 2, 85748 Garching bei M\"unchen, Germany\\
$^6$Instituto de Astrofisica de Canarias, 38205 La Laguna, Tenerife, Spain\\             
$^7$Kapteyn Astronomical Institute, University of Groningen, Postbus 800, 9700 AV, Groningen, The Netherlands\\
$^8$Leiden Observatory, Leiden University, Postbus 9513, 2300 RA, Leiden, The Netherlands\\
$^9$Max Planck Institut f\"ur extraterrestrische Physik, Giessenbachstrasse, 85748 Garching, Germany\\
$^{10}$Universit\'e de Strasbourg, CNRS, UMR 7550, F-67000 Strasbourg, France\\ 
$^{11}$Max-Planck-Institut f\"ur Astronomie, K\"onigstuhl 17, D-69117 Heidelberg, Germany\\
$^{12}$INAF - Osservatorio Astronomico di Roma, via Frascati 33, 00040 Monteporzio, Italy\\ 
$^{13}$Space Telescope Science Institute, Baltimore, MD 21218 \\           
$^{14}$ Universidad de La Laguna, Dpto. Astrofisica, E-38206 La Laguna, Tenerife, Spain\\
$^{15}$Dipartimento di Fisica \& Astronomia, Universit\`a degli Studi di Bologna, Viale Berti Pichat, 6/2, I - 40127 Bologna, Italy\\
}

\date{Accepted for publication on February 15, 2018}

\pubyear{2018}

\begin{document}
\label{firstpage}
\pagerange{\pageref{firstpage}--\pageref{lastpage}}
\maketitle

\begin{abstract}
SECCO~1 is an extremely dark, low-mass ($M_{\star}\simeq 10^5$~M$_{\sun}$), star-forming stellar system lying in the Low Velocity Cloud (LVC) substructure of the Virgo cluster of galaxies, and hosting several H{\sc ii} regions. Here we review our knowledge of this remarkable system, and present the results of (a) additional analysis of our panoramic spectroscopic observations with MUSE, (b) the combined analysis of Hubble Space Telescope and MUSE data, and (c) new narrow-band observations obtained with OSIRIS@GTC to search for additional H{\sc ii} regions in the surroundings of the system. We provide new evidence supporting an age as young as $\la 4$~Myr for the stars that are currently ionising the gas in SECCO~1. We identify only one new promising candidate H{\sc ii} region possibly associated with SECCO~1, thus confirming the extreme isolation of the system. We also identify three additional candidate pressure-supported dark clouds in Virgo among the targets of the SECCO survey. Various possible hypotheses for the nature and origin of SECCO~1 are considered and discussed, also with the help of dedicated hydrodynamical simulations showing that a hydrogen cloud with the characteristics of SECCO~1 can likely survive for $\ga 1$~Gyr while traveling within the LVC Intra Cluster Medium.
\end{abstract}




\section{Introduction}

SECCO~1 is a faint star-forming stellar system that was discovered by 
\citet[][B15a and B15b, hereafter]{pap1,secco1} in a survey (SECCO\footnote{\tt http://www.bo.astro.it/secco}; B15a) aimed at 
searching for stellar counterparts of Ultra Compact High Velocity H{\sc i} clouds (UCHVC), selected from the ALFALFA\citep{adams} radio survey.

In B15b we proved the physical association of the small group of blue compact sources originally identified in B15a with the UCHVC HVC~274.68+74.0, from the sample of \citet[][]{adams}, and we obtained a first estimate of the oxygen abundance, that appeared anomalously high for a dwarf galaxy as faint as SECCO~1. 
Subsequently \citet[][S15, hereafter]{sand15} reported on the independent discovery of the same system, confirming the results of B15b. These authors identified a smaller system with similar appearance lying just $\sim 2\arcmin$ apart \citep[see also][Be16, hereafter]{pap2}. In the following we will refer to the two pieces of SECCO~1 as the Main Body (MB) and the Secondary Body (SB), according to \citet[][Be17a, hereafter]{pap_muse}. Both B15b and S15 agree that SECCO~1 lies within the Virgo cluster of galaxies. In particular, it probably belongs to a substructure of the cluster known as low-velocity cloud \citep[LVC,][]{boselli}, whose central galaxy is the spiral NGC~4216. The Hubble Space Telescope photometry of SECCO~1 obtained by \citet[][S17, hereafter]{sand17} is consistent with this conclusion.

High spatial resolution H{\sc i} observations by \citet[][A15, hereafter]{adams15} split the original HVC~274.68+74.0 into two smaller clouds: AGC~226067, centered on MB, and AGC 229490, off-set by $\simeq 0.5 \arcmin$ with respect to SB.
Assuming a distance D=17.0~Mpc for the LVC \citep[following][]{boselli}, the total H{\sc i} mass associated to SECCO~1 is 
$M_{HI}\sim 2\times 10^7 M_{\sun}$. On the other hand, the total stellar mass derived with two independent methods from HST data by Be17a and S17 is $M_{\star}\la 1.6 \times 10^5 M_{\sun}$, thus implying a H{\sc i}-to-stellar mass ratio of $\frac{M_{HI}}{M_{\star}}\ga 100$, fully in the range of almost-dark galaxies \citep{can_dark}\footnote{Note that S17 provide an $M_V$ estimate that is 0.9 mag brighter than that by B17a, but, given the large uncertainties, this corresponds to a $2\sigma$ difference. We feel more confident on the B17a estimate since there, while summing the light of SECCO~1, we were able to exclude the contribution from unrelated sources that we unambiguously identified, based on their velocity, thanks to the MUSE spectra.}. 

Finally, in Be17a, we presented the results of panoramic optical spectroscopy of SECCO1 MB and SB  obtained with the integral-field spectrograph MUSE@VLT \citep{bac14}. The MUSE data-cube allowed us to identify many individual H{\sc ii} regions in both MB and SB, measuring their radial velocity and metallicity. We found that MB and SB not only have very similar systemic velocities but also the same oxygen abundance, thus demonstrating that the two systems have a common origin. 

In particular, all the H{\sc ii} regions in which we could obtain a reliable estimate of the oxygen abundance have the same abundance (from strong-lines indicators) within the observational uncertainties. 
The mean metallicity is $\langle 12+(O/H)\rangle=8.38 \pm 0.10$, confirming the earlier estimate by B15b, a value typical of galaxies as massive as the Large Magellanic Cloud or M33. The anomalously high metallicity, for the small stellar mass, strongly suggests that the gas that is presently converted into stars in SECCO~1 was chemically enriched in a much larger galaxy, from which it was torn apart by tidal or ram-pressure interaction with the harsh Virgo environment. The apparent lack of a stellar population older than a few tens of Myr 
provides additional support to this hypothesis (Be17a, S17).
The star formation was (independently) found by Be17a,b\footnote{The SFR originally reported in B17a was flawed by a trivial computation error that has been corrected in \citet{err_muse}.} and by S17  to have a rate typical of star-forming dwarfs with total luminosity comparable to SECCO~1 \citep[][see Table~\ref{mean}, below]{bdd}.

While star formation is known to occur in ram-pressure stripped gas clouds in galaxy clusters \citep[see, e.g.,][and references therein]{gerhard,yoshi,kenney,hester,fuma,fumaram,fossa} and in the surrounding of interacting galaxies \citep[see, e.g.,][]{RW04}, this usually happens in relatively close proximity to the stripped galaxies. SECCO~1 is by far the most isolated case ever observed (S17, Be17a), as the closer plausible progenitor proposed up to now (the M86 subgroup, S17) lies at more than 350 kpc (but see below). This implies that, if indeed SECCO~1 is a gas cloud stripped from that group, it must have travelled within the hot Intra Cluster Medium (ICM) for about 1~Gyr before the ignition of the current star formation episode (Be17a, S17). Hence SECCO~1 is of special interest to understand the behaviour of such long-lived quiescent stripped clouds that may be present in large numbers within galaxy clusters \citep{burk}.

In Table~\ref{mean} we provide a summary of the main physical properties of SECCO~1, adopting the distance $D=17.0$~Mpc, following A15, Be17a and S17. This is the distance to SECCO~1 that we always adopt in this paper.

In Be17a we presented the observational framework for this intriguing stellar system. Here we discuss and investigate in detail its origin and evolution, with the support of dedicated hydrodynamical simulations. 
Sect.~\ref{datared} is devoted to describe the reduction of the HST images of SECCO~1 already presented by S17, and of the deep narrow-band H${\alpha}$ images that we obtained with OSIRIS@GRANTECAN (GTC).
In Sect.~\ref{port} we present additional results from the MUSE data, in particular on the effect of the ionisation on the estimate of metallicity and on the nature of ionising sources. In Sect~\ref{cmd} we analyse the stellar content of SECCO~1, coupling the constraints from HST and MUSE data. In Sect.~\ref{pieces} we present the results of the GTC H${\alpha}$ imaging. In Sect.~\ref{disc} we put SECCO~1 in the context of dwarf galaxies and of stripped gas clouds, discussing the pros and cons of various evolutionary paths. In Sect.~\ref{simu} we present the results of 2D and 3D hydrodynamical simulations aimed at establishing the lifetime and evolution of a pressure-supported H{\sc i} gas cloud similar to SECCO~1 travelling into the ICM. In Sect.~\ref{starf} we discuss the conditions for the occurrence of star formation in H{\sc i} clouds similar to SECCO~1 and we identify three new candidate dark gas clouds that may be wandering within the Virgo ICM like SECCO~1. Finally, in Sect.~\ref{conc} we summarise our results.  

\section{Data Analysis}
\label{datared}

In this section we briefly report on the reduction of the HST images of SECCO~1 obtained by S17, that will be discussed in Sect.~\ref{cmd}, and of the acquisition and reduction of narrow-band $H{\alpha}$ imaging, whose results are presented in Sect.~\ref{pieces}.

\subsection{HST photometry}
\label{datared_hst}
S17 present a detailed analysis of Hubble Space Telescope (HST) Advanced Camera for Surveys (ACS) Wide Field Camera (WFC) F606W and F814W images of SECCO~1 (GO~13735 program, P.I.: D.J. Sand).

We reduced and analysed independently the S17 dataset to have a deeper insight on the stellar populations in SECCO~1 by combining the HST-ACS and MUSE data. 
Positions and photometry of individual stars were obtained with the ACS module of the point spread function (PSF) fitting package DOLPHOT v.2.0 \citet{dolph}, as described, for example, in \citet{vv124acs}. We adopted a threshold of 2.5$\sigma$ above the background for the source detection. Photometry was performed on the individual images corrected for charge transfer efficiency (FLC files), adopting the F814W drizzled distortion-corrected image (DRC file) as the reference for source identification.
In the following we will use magnitudes in the ACS-WFC VEGAMAG system and the analysis will be limited to the best quality sources, selected by  having quality flag=1, sharpness parameter $\lvert {\tt sharp}\rvert<0.2$ for F814W$\le 26.5$ and $-0.4<{\tt sharp}<0.2$ for F814W$> 26.5$, crowding parameter ${\tt crowd} < 1.0$, goodness of fit parameter ${\tt chi} <2.0$, and photometric errors in both passbands $\le 0.5$~mag. The resulting CMD is discussed in Sect.~\ref{cmd}.

\subsection{GTC $H{\alpha}$ imaging}
\label{datared_gtc}

S17 report that there is no evidence of further pieces of SECCO~1 within the field sampled by their ACS images.
However the ACS-WFC field is relatively small and the most interesting (and easy to find) kind of sources to be searched for in the surroundings of SECCO~1 are H{\sc ii} regions, isolated or in groups, that can be associated to the system. Indeed, e.g., ram pressure stripping episodes are expected to produce star-forming blobs in large numbers \citep[see][and the discussion in Sect.~\ref{disc}, below]{kap09}. 

To look for these sources we obtained H${\alpha}$ imaging of a $\simeq 7.8\arcmin \times 7.8\arcmin$ field centered on SECCO1 MB with the OSIRIS\footnote{\tt http://www.gtc.iac.es/instruments/osiris/osiris.php} camera mounted on the 10.4m Gran Telescopio CANARIAS (GTC), at the Observatory Roque de los Muchachos, in La Palma, Spain (program GTC42-16A, P.I.: G. Battaglia).
We used the OSIRIS tunable filters \citep{gonza} f657/35, sampling H${\alpha}$ rest-frame (FWHM$\simeq 35$~nm), and f680/43, sampling an adjacent portion of the continuum (FWHM$\simeq 43.2$~nm). 
We obtained a total of twenty-one $t_{exp}=600$~s images per filter, applying a dithering of a few arcsec between each individual image, in various service mode visits from January to March 2017.
The typical seeing was around $\sim 1.0\arcsec$.

All the images were corrected for flat-field and bias and the sky was subtracted, also removing the sky ring patterns that are typical of this observing mode of OSIRIS \citep{gonza}. Then all the images were combined into one single stacked f657/35 image and one single stacked f680/42 image. We used Sextractor \citep{sex} to identify sources with peak intensity $\ga 10\sigma$ above the backgroundand to measure their magnitudes (mag$_{H{\alpha}}$ and mag$_{cont}$) in both images. The output catalogs were combined using codes from the Catapack 
suite\footnote{By  P. Montegriffo, \tt http://davide2.bo.astro.it/\~\ paolo/Main/CataPack.html}. Then, in a plot 
mag$_{H{\alpha}}$ vs. color (mag$_{H{\alpha}}$ - mag$_{cont}$), we identified the sources having an excess in mag$_{H{\alpha}}$ - mag$_{cont}$  larger than three times the standard deviation in a broad range around their H${\alpha}$ luminosity. In this way we obtained a catalog of a handful of candidate H${\alpha}$ emitters, including already-known SECCO~1 sources as well as several sources which are detected only in H${\alpha}$ and not in continuum images. The analysis of this catalog is described in Sect.~\ref{pieces}.

\begin{table*}
  \begin{center}
  \caption{Physical properties of SECCO~1 (assuming D=17.0~Mpc)}
  \label{mean}
  \begin{tabular}{lcccc}
\hline
parameter       & MB  & SB & Total & note\\
\hline
RA$_{J2000}$      &12:21:54.0  & 12:21:55.7&  &  from the ACS-WFC F606W image$^a$\\
Dec$_{J2000}$     &+13:27:36.8 & +13:29:02.3& &  from the ACS-WFC F606W image$^a$\\
L$_V$[$L_{V,\sun}]$   & $1.2^{+0.5}_{-0.4}\times 10^{6}$ &$4.4^{+2.6}_{-1.6}\times 10^{5}$&$1.6^{+0.6}_{-0.4}\times 10^{6}$ & from Be17a\\
M$_{\star}$[M$_{\sun}$] &$\la 1.2\times 10^{5}$ & $\la 0.4\times 10^{5}$&$\la 1.6\times 10^{5}$ &upper limit from integrated light (Be17a)\\
M$_{\star}$[M$_{\sun}$] &$7.3^{+3.0}_{-2.4}\times 10^{4}$ &$2.7^{+1.6}_{-1.0}\times 10^{4}$&$1.0^{+0.4}_{-0.3}\times 10^{5}$ &from integrated light (Be17a)$^{b}$\\
M$_{\star}$[M$_{\sun}$] &$5.4\pm 1.3\times 10^{4}$ & & &from CMD analysis (S17)\\
M$_{HI}$[M$_{\sun}$] &$1.5\times 10^7 $&$3.6\times 10^6$ &$1.9\times 10^7$ & from A15$^c$\\
Radial size$_{\star}$ [kpc] & 1.2 & 1.2 & &radius enclosing all the H${\alpha}$ emission (Be17a)\\
r$_{HI}$ [kpc] & 3.7 & $<1.6$ & &H{\sc i} radius from A15\\
$V_h^{\star}$ [km~s$^{-1}$] & $-153.2\pm 1.4$ &$-126.5\pm 2.5$ & &heliocentric radial velocity (stars, Be17a)\\
$V_h^{HI}$ [km~s$^{-1}$]& $-142$ & $-123$ & &heliocentric radial velocity (H{\sc i}, A15)\\
$\sigma_{V_h}^{\star}$ [km~s$^{-1}$]& $3.5\pm 2.1$ & $2.7\pm 5$ & & from Be17a\\
$\sigma_{V_h}^{HI}$ [km~s$^{-1}$]& $9\pm 3$ & $4\pm 2$ & & from A15\\
$\langle12+log(O/H)\rangle$ & $8.37\pm 0.11$ &$8.39\pm 0.11$  & $8.38\pm 0.11$ & from Be17a,b\\
SFR [M$_{\sun}$~yr$^{-1}$]& & &$0.7\pm 0.2\times 10^{-3}$ &from H${\alpha}$ flux (Be17a,b)$^d$\\
SFR [M$_{\sun}$~yr$^{-1}$] & & &$1.1\times 10^{-3}$ &from CMD analysis (S17)\\
D$_p$(MB-SB) [kpc] & & & 7.3 &proj. distance between MB and SB\\
D$_p$(stars-H{\sc i}) [kpc] & $\simeq 0.0$ &$\simeq 2.5$ & &  proj. distance between stars and gas cloud\\
D$_p$(MB-NGC4299) [kpc] & 582   &  & & proj. distance from proposed parent system (Be17a)\\
D$_p$(MB-M86) [kpc] & 346   &  & & proj. distance from proposed parent system (S17)\\
D$_p$(MB-IC3142) [kpc] & 255   &  & & proj. distance from proposed parent system (this work)\\
\hline	      
\multicolumn{5}{l}{$^a$ Geometric centers of the distribution of sources, estimated by eye. For MB this is slightly different than that adopted in B17a that}\\ 
\multicolumn{5}{l}{referred to the center of the main northern clump of sources. To establish the geometric center of SB we used the distribution of H{\sc ii}}\\ 
\multicolumn{5}{l}{regions on the MUSE datacube as a guideline.}\\
\multicolumn{5}{l}{$^b$ adopting M/L$_V$=0.061 from the solar-scaled BASTI model \citep{basti2} with age=30~Myr and [Fe/H]=-0.66.}\\ 
\multicolumn{5}{l}{$^c$ The H{\sc ii} AGC229490 cloud is off-set by 0.5$\arcmin$ from SB.}\\
\multicolumn{5}{l}{$^d$ The reported uncertainty combines the error on the integrated H${\alpha}$ flux and the uncertainty associated to the adopted}\\ 
\multicolumn{5}{l}{calibration \citep{kennicutt}. Varying the assumed distance by $\pm 3.0$~Mpc changes the SFR by $\simeq \pm 0.2\times 10^{-3}$~M$_{\sun}$~yr$^{-1}$.}
\end{tabular} 
\end{center}
\end{table*}

   \begin{figure}
   \includegraphics[width=\columnwidth]{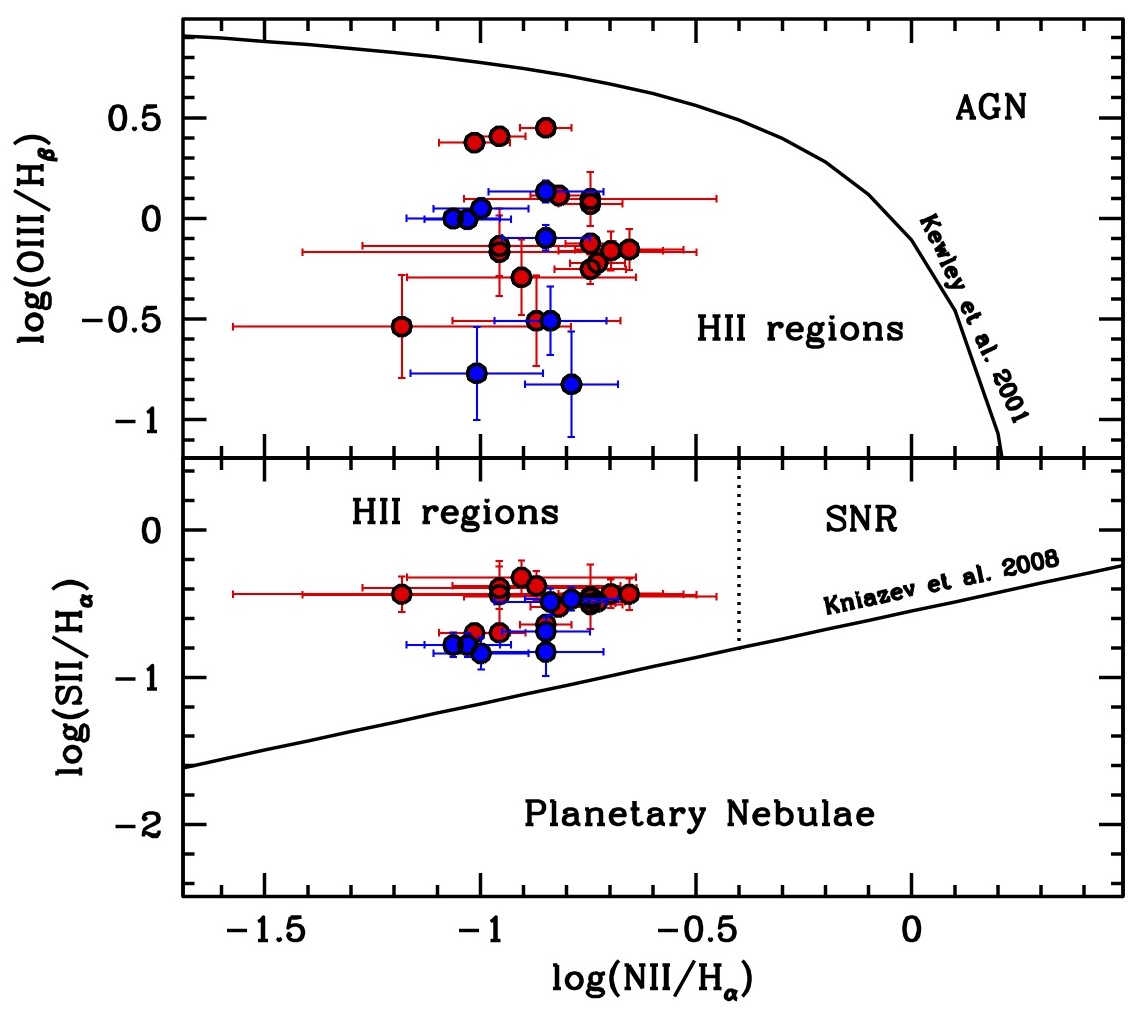}
     \caption{Classification diagnostic diagrams for all the SECCO~1 sources identified in Be17a for which reliable estimate of the flux for the relevant lines could be obtained, following \citet{kniazev}. Red filled circles are sources residing in MB, blue filled circles lie in SB. The threshold curves are from \citet[][upper panel]{kewley}, and from \citet[][lower panel]{kniazev}.
     The dotted line in the lower panel marks the lower limit in log(N{\sc ii}/H${\alpha}$) of the locus occupied by Supernova Remnants. The upper(lower) panel includes 24(23) of the 38 sources identified by Be17a.} 
        \label{ratios}
    \end{figure}

\section{More from MUSE: ionisation, kinematics and ionising sources}
\label{port}

As a first step in our review of the properties of SECCO~1, we present some additional analysis of the MUSE data presented in Be17a (see this paper for all the details on the observational material and data reduction).
In Fig.~\ref{ratios} we show the position of the individual sources identified in Be17a in the diagnostic diagrams that are generally used to discriminate H{\sc ii} regions from Active Galactic Nuclei (AGN), Supernova remnants (SNR) and Planetary Nebulae, based on line flux 
ratios. According to these diagnostic diagrams, all the considered sources are unambiguously classified as H{\sc ii} regions.

Given the star formation history inferred by S17 we expect that at least some massive stars (conventionally those with $m_i\ge 8.0$~M$_{\sun}$) have already exploded as type II Supernova (SNII). 
We use the fraction of the total mass contributed by massive stars in the stellar mass ranges $8.0$--$40$ ~M$_{\sun}$ and $40$--$100$~M$_{\sun}$, computed by \citet{romano} for several different 
Initial Mass Functions \citep[IMFs][]{salp, tinsley80, scalo86, scalo98, kroupa93, chabrier03}, to estimate the quantity of SNe exploded in SECCO~1, adopting $M_{\star}=1.0\times 10^5~M_{\sun}$ from Tab.~\ref{mean}. The mass of stars with $m=8-40~M_{\sun}$ ranges from a a minimum of about $5500~M_{\sun}$ with the IMF of \citet{scalo86} to a maximum of 1.8$\times$10$^4~M_{\sun}$ with the IMF of \citet{tinsley80}. Stars in the range $m=40-100~M_{\sun}$ contribute about 300~M$_{\odot}$ \citep{tinsley80} to $\sim6000~M_{\sun}$ \citep{chabrier03} to the total stellar mass budget of the system.  Since, according to the \citet{parsec} stellar models of the proper metallicity (Z=0.006), the lifetime of a 40~$M_{\sun}$ star is $\simeq 5.0$~Myr and that of a 8~$M_{\sun}$ star is $\simeq 40.0$~Myr, we can conclude that, if the IMF of SECCO~1 is comprised within the range of cases considered here and independently of the detailed star formation history, several SNII explosions should have occurred in this system.
Is this consistent with the fact that we do not see signatures of shocks in the spectra of the ionised regions of SECCO~1? Considering that  in the spiral galaxy M33, \citet{asvarov14} computed a number of about  1000 expected  {\em alive} (i.e., observable) Supernova remnants (SNR), we can just scale 
that number by the stellar mass ratios of the two galaxies ($\sim$10$^{-4}$), to roughly estimate the number of expected SNRs in SECCO1, obtaining only 0.1 SNRs. This number is indeed compatible with the non detection of SNRs in our observations. 


\begin{figure*}
   \includegraphics[width=0.8\textwidth]{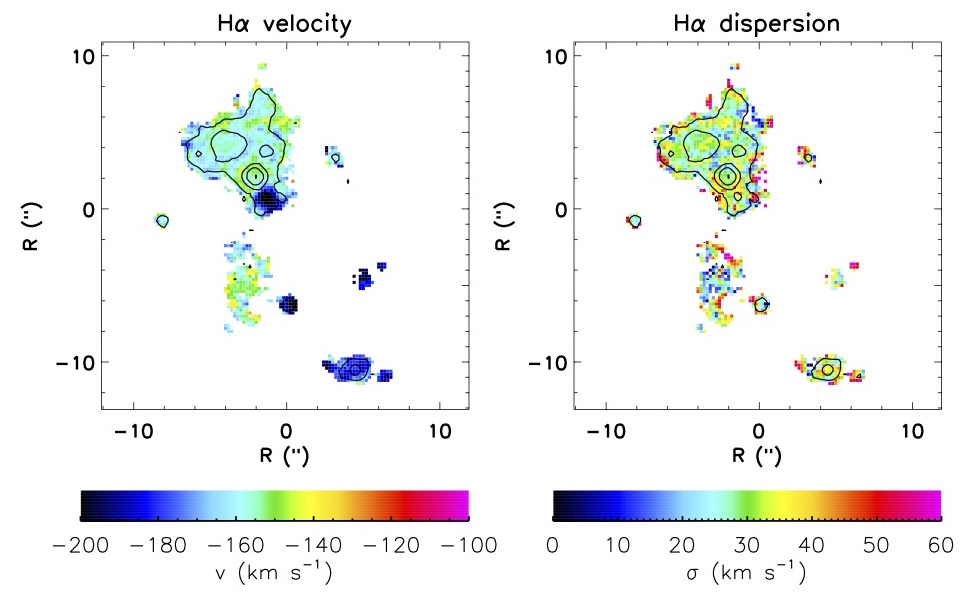}
     \caption{Kinematic maps for the Main Body. The H$\alpha$ velocity (left panel) and velocity dispersion (right panel) are shown for all the spaxels with $S/N>3$ after a Gaussian kernel smoothing of 3 pixels ($0.6\arcsec$). H$\alpha$ contours are overplotted for reference, corresponding to 0.3, 0.8, 2, 6 erg\ s$^{-1}$\ cm$^{-2}$\ \AA$^{-1}$ respectively.} 
        \label{hadynmb}
\end{figure*}

\begin{figure*}
   \includegraphics[width=0.8\textwidth]{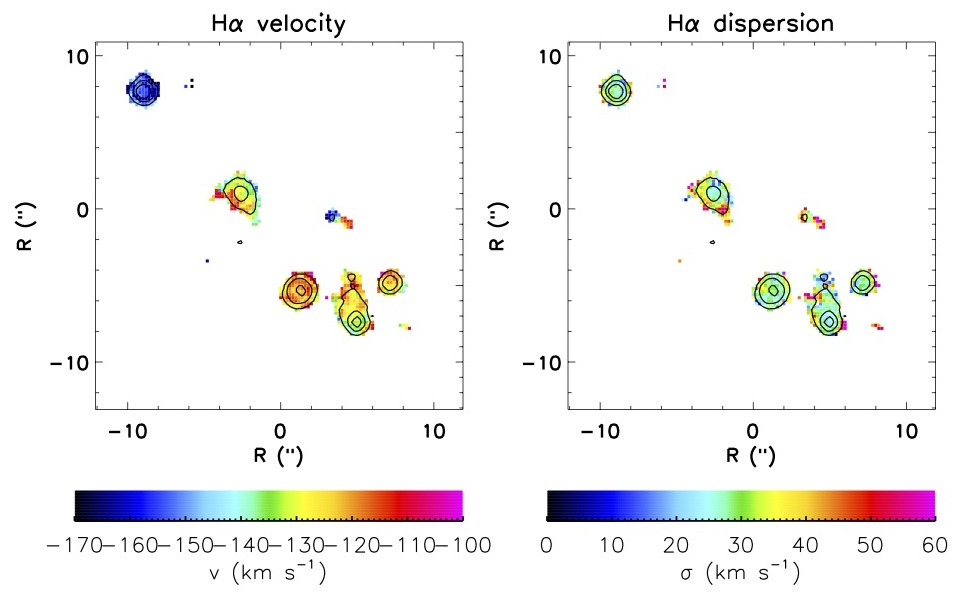}
     \caption{Kinematic maps and  H$\alpha$ contours for the Secondary Body. 
     The arrangement and the meaning of the symbols are the same  as in Fig.~\ref{hadynmb}.} 
        \label{hadynsb}
\end{figure*}

\begin{figure*}
\includegraphics[width=1.0\textwidth]{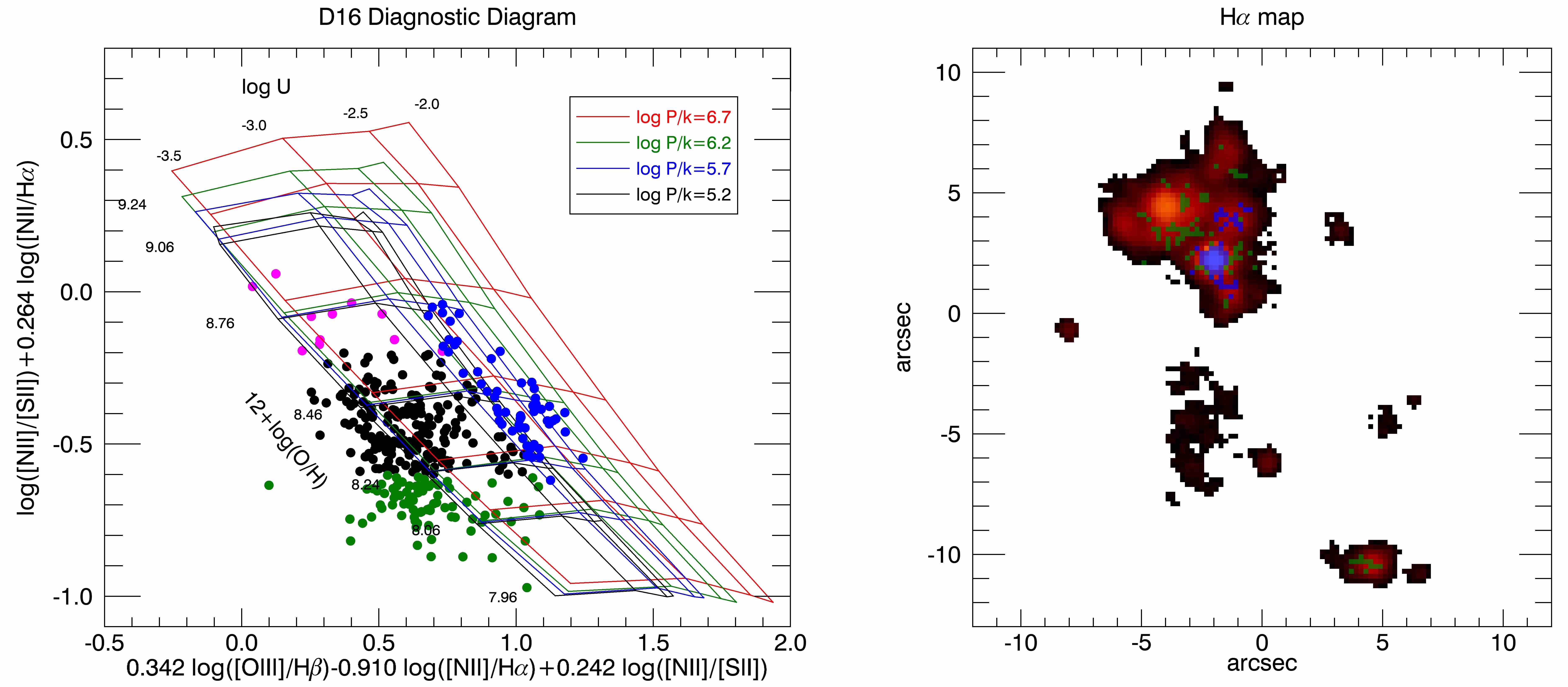}
     \caption{The spatially resolved emission line diagnostic by Dopita et al. (2016) for spaxels in the SECCO Main Body with S/N$>3.0$ in all the emission lines involved. The diagnostic diagram is shown in the left panel, where each point correspond to a MUSE spaxel. The metallicity scale (increasing towards the upper part of the plot) and ionisation parameter scale (increasing towards the right part) are shown on the different grids, corresponding to different values of the ISM pressure. The spaxels whose metallicity in the Dopita et al. (2016) scale is 12 + log(O/H) $<8.25$ are plotted as green filled circles, those with 12 + log(O/H) $>8.6$ as magenta filled circles, while the spaxels with log~U$>-3.0$ are shown as blue filled circles, independently of their oxygen abundance. The same spaxels are over-plotted with the same color code on the H$\alpha$ map in the right panel.} 
        \label{dopitamb}
\end{figure*}

\subsection{Kinematics, metallicity and ionisation of the diffuse gas}

In Fig.~\ref{hadynmb} and Fig.~\ref{hadynsb} we show the H$\alpha$ velocity map (left panel) and velocity dispersion map (right panel), corrected for the intrinsic instrumental broadening for MB and SB, respectively. 
The maps show only the spaxels with $S/N>3$ in the H$\alpha$ line, whose parameters are derived by fitting a single Gaussian to every spaxel. 

This view is complementary to the analysis of the individual H{\sc ii} regions presented in Be17a, as it traces also the properties of diffuse ionised gas that is present in SECCO~1.
Not surprisingly, the systemic velocity of the diffuse gas follows the same velocity field as individual sources. The velocity dispersion of the gas within the nebulae (not shown in Be17a) is very uniform over the body of the system, with mean values typical of H{\sc ii} regions. 

As discussed in Be17a (and already noted in B15b), some of the sources (in particular MB25, MB59 and MB61) show much stronger [OIII] emission than the others (specifically, higher [OIII]/H${\beta}$ ratios). This may be due to a different metallicity or to a different degree of ionisation. To investigate the actual origin of the anomalous [OIII] emission in these regions we use the diagnostic diagram proposed by \citet[][D16 hereafter]{dop16}, that provides an estimate of both the ionisation parameter U and the oxygen abundance 12+log(O/H) from a combination of rest-frame optical strong line ratios. 

The D16 diagram is shown in the left panel of Fig.~\ref{dopitamb}, where the different grids shows the variation of metallicity and log~U for different values of ISM pressure. The spaxels with $S/N>3.0$ in all the lines involved ([OIII], H$\alpha$, H$\beta$, [NII] and [SII] doublet) are shown as points on the grid, colored according to their position in the diagram (outliers in log~U in blue, spaxels in the high-metallicity tail in magenta, those in the low metallicity tail in green; all the others in black). 
The position of the blue, green and magenta spaxels is superposed to the H${\alpha}$ emission map in the right panel of Fig.~\ref{dopitamb}. The high-ionisation blue points are all clustered at the position of the sources with the anomalosuly high [OIII] emission, strongly suggesting that the larger [OIII]/H${\beta}$ ratio is not due to variation in metallicity, but to a higher ionisation parameter at these locations.
The spaxels tracing metallicities lower than 12+log(O/H)$<8.25$ (green points) appear to correspond to diffuse gas between the detected H{\sc ii} regions, where the mixing of metals diffused by stellar winds and supernova explosions within each nebula is probably less effective. In any case, the metallicity variations are $\pm0.25$ dex across the whole system, comparable to the uncertainties. Finally, if we limit the analysis to the spaxels with $S/N>5.0$ in all the involved lines, {\em all} the magenta dots and most of the green ones (in particular, {\em all} those having 12+log(O/H)$\le 8.06$) disappear, showing that the tails of the metallicity distribution are populated by the noisier spaxels\footnote{Note that in Be17a the contribution of all the spaxels associated to a given source were integrated together, providing source spectra with much higher S/N and, consequently, more robust metallicity estimates for the individual H{\sc ii} regions.}. The high degree of chemical homogeneity of the system found in Be17a is confirmed using metallicity estimates from the D16 diagram. Fully analogous conclusions hold for SB. We do not show the D16 diagram for SB because in that piece of SECCO~1 there is no source displaying an anomalously high [OIII] emission.   


\begin{table*}
\caption{Temperature and Spectral Type of ionising stars}
\label{Tab_flux}
\begin{tabular}{|l|c|c|c|c|c|}
\hline
\multicolumn{1}{|l|}{Name} &
\multicolumn{1}{c|}{He I 5786} &
\multicolumn{1}{c|}{H$\beta$} &
\multicolumn{1}{c|}{  $\frac{He I}{H\beta}$ } &
\multicolumn{1}{c|}{$T_{eff}$} &
\multicolumn{1}{c|}{ST$^a$} \\
\hline
unit       &  erg~cm$^{-2}$s$^{-1}$& erg~cm$^{-2}$s$^{-1}$& & K & \\
\hline
MB25 & 7.1$\pm$1.7E-18 & 1.07$\pm$0.02E-16 & 0.07$\pm$0.02 & 35000$\pm$2000 & O7.5V\\
MB26 & 8.1$\pm$2.3E-18 & 5.00$\pm$0.02E-17 & 0.16$\pm$0.05 & 53000$\pm$17000$^{b,c}$ & O2V\\
MB27 & 1.1$\pm$2.6E-17 & 9.13$\pm$0.02E-17 & 0.12$\pm$0.03 & 40000$\pm$3000 & O5.5V\\
MB30 & 9.5$\pm$2.1E-18 & 6.86$\pm$0.02E-17 & 0.14$\pm$0.04 & 47000$\pm$10000$^c$ & O2V\\
MB56 & 9.3$\pm$2.2E-18 & 8.00$\pm$0.02E-17 & 0.12$\pm$0.03 & 40000$\pm$3000 & O5.5V\\
MB61 & 1.2$\pm$0.2E-17 & 8.65$\pm$0.02E-17 & 0.14$\pm$0.03 & 42000$\pm$2000 & O4V\\
SB2 & 5.2$\pm$1.6E-18 & 6.21$\pm$0.02E-17 &    0.08$\pm$0.03  & 36000$\pm$2000 & O7V\\
SB45 & 5.4$\pm$1.8E-18 & 6.08$\pm$0.02E-17 & 0.09$\pm$0.03 & 37000$\pm$3000 & O6.5V\\
\hline
\multicolumn{6}{l}{$^a$ Spectral types are from \citet{PM13} for $T_{eff} \le$46000~K, while we associate the spectral type O2V}\\ 
\multicolumn{6}{l}{ to stars with $T_{eff} >$46000~K.}\\
\multicolumn{6}{l}{$^b$ $T_{eff}$ of this star is extrapolated from the relation given in Fig.~9 of 
\citet[][lower continuous curve]{K00},}\\ 
\multicolumn{6}{l}{since the relation is valid only for $T_{eff} <$50000.}\\
\multicolumn{6}{l}{$^c$ The large errors on the $T_{eff}$ for these stars are due to the flattening of the relation between  $\frac{He I}{H\beta}$ and $T_{eff}$ for $\frac{He I}{H\beta}>$ 0.12.}
\end{tabular}
\end{table*}

   \begin{figure*}
   \includegraphics[width=\textwidth]{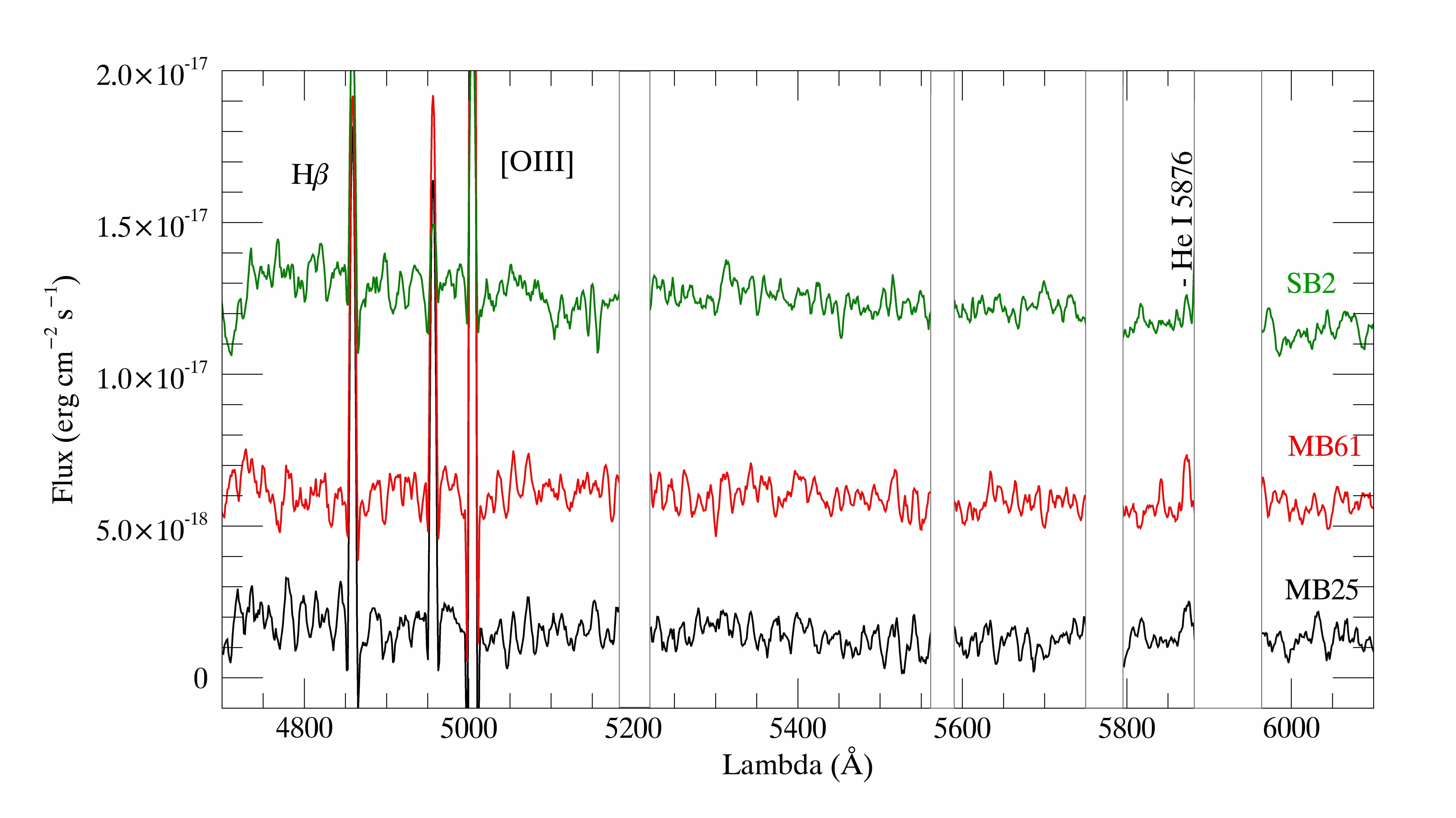}
     \caption{MUSE spectra (from B17a) of three H{\sc ii} regions listed in Tab.~\ref{Tab_flux}, representative of the quality of the whole sample (in terms of flux in the He{\sc I}5786 line). The spectra of the sources with the highest and the lowest He{\sc I}5786 fluxes (MB61 and SB2, respectively) are plotted, as well as an intermediate case (MB25). The spectral range shown include the lines used to estimate the temperature of the ionising stars. Arbitrary shifts in flux are applied to the spectra of MB61 and SB2 to make the plot readable. Blank vertical bands mask regions of the spectra badly affected by the residual of strong sky lines. } 
        \label{spectrum}
    \end{figure*}

\subsection{Spectroscopic constraints on the age of the youngest stellar population}
\label{laura}

To ionise an H{\sc ii} region we need O or B stars. \citet{D17} found that most extragalactic H{\sc ii} regions are ionised by stars in the $T_{\rm eff}$ range 37-40~kK, with an average value of 38.5~kK, corresponding to a MS lifetime shorter than 4 Myr \citep{PM13}. This suggests that also in SECCO~1 the ionising stars can belong to a very young population that may be difficult to unambiguously identify and characterise in a sparsely populated CMD. Note that 4~Myr is the age of the youngest isocrone in the PARSEC\footnote{\tt http://stev.oapd.inaf.it/cgi-bin/cmd} set \citep{parsec} that we use in Sect.~\ref{cmd}. 
To investigate this issue we attempt to constrain the temperature and, 
consequently, the age of the stars powering the HII nebulae in SECCO~1, using our MUSE spectra.   

The effective temperature ($T_{\rm eff}$) of the ionising stars of a nearby H{\sc ii} region can be estimated through  direct observation and spectral classification  \citep[see e.g.][]{walborn14,  evans15, lamb15}. 
For more distant stars, we have an indirect estimate of $T_{\rm eff}$ based on the analysis of the emission lines of the nebular gas as proposed originally by \citet{zanstra31} and later by  \citet{os06}.
In particular, we have used the ratio between the He{\sc i} 5876\AA\ line and H$\beta$ which gives an estimate of the effective temperature nearly independent of the ionising parameters \citep[see, e.g., Fig.~9 of][]{K00}.

   \begin{figure*}
   \includegraphics[width=0.8\textwidth]{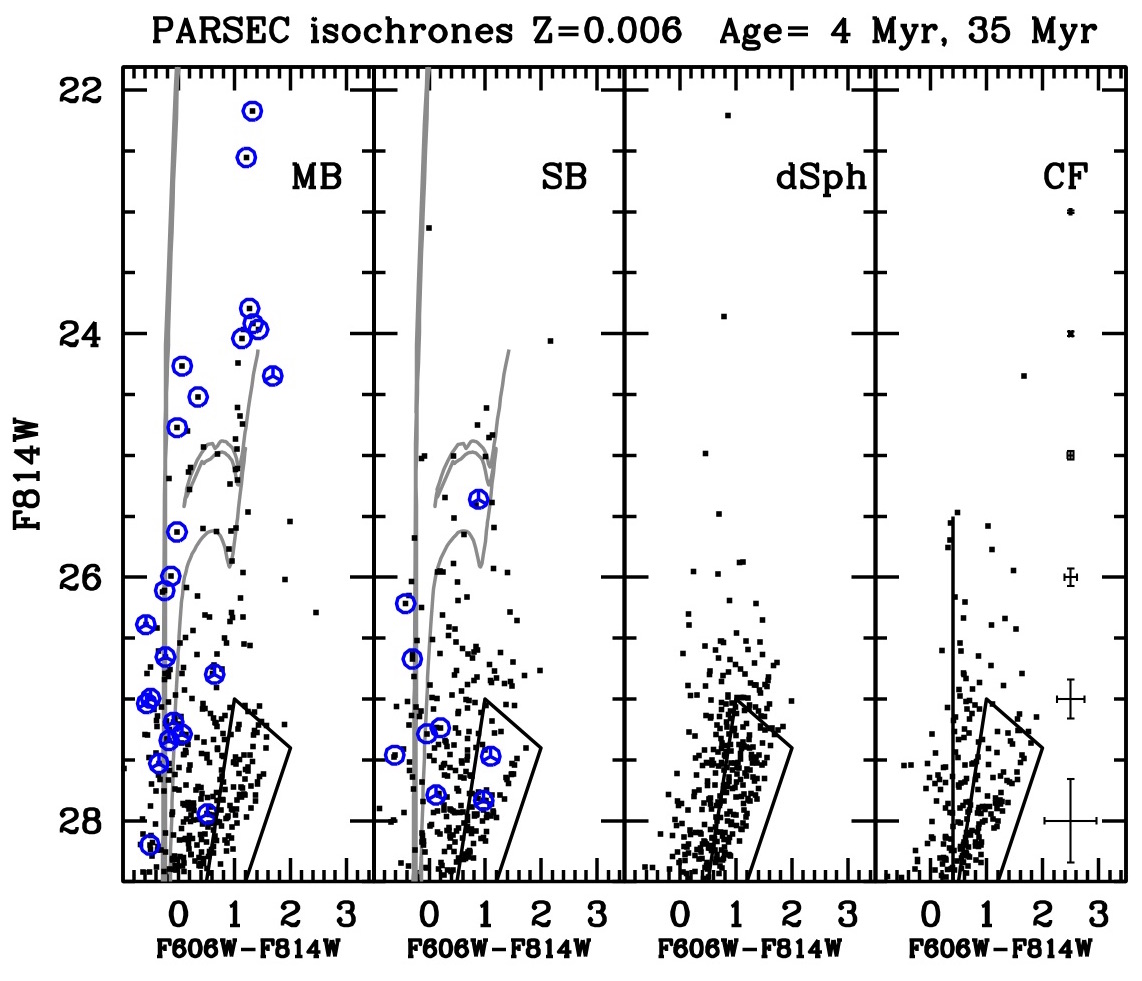}
     \caption{Color magnitude diagrams from HST/ACS photometry of circular regions of radius=$20\arcsec$ centered, from left to right, on the center of the main body of SECCO~1, on the center of the secondary body of SECCO~1, on the center of the Virgo dwarf spheroidal Dw~J122147+132853, and on a region devoid of resolved stellar systems (Control Field). The polygonal contour at F814W$\le 27.0$ and F606W-F814W$\simeq 1.0$ approximately encloses the intra cluster population of RGB stars as sampled by \citet{williams}, the thick vertical segment at F606W-F814W$=0.4$ is the ridge line of the blue intra cluster population from the same source (the difference in E(B-V) between the two lines of sights is $<0.03$ mag and it has been neglected). 
Blue circles are sources cross-identified in the list of H{\sc ii} regions in B17a; those having a skeletal triangle inscribed are recognised as more extended than individual point-sources by DOLPHOT. To superpose the PARSEC isochrones (grey curves) we adopted D=17~Mpc and $E(B-V)=0.048$.}
        \label{cocm}
    \end{figure*}

In Tab.~\ref{Tab_flux} we summarise our results for all the B17a H{\sc ii} regions where the He{\sc i} line can be reliably measured from our MUSE data. In Fig.~\ref{spectrum} we show the relevant portion of the MUSE spectrum for a representative sample of the sources listed in Tab.~\ref{Tab_flux}. The table reports the observed fluxes of the H$\beta$ and of the He{\sc i} lines, their ratios, the corresponding $T_{\rm eff}$, and an indicative spectral classification. 
All stars belong to the O spectral type, being all hotter than $>$ 35~kK. This means that they are more massive than 20 $M_{\sun}$ and indeed younger than $\simeq 4$~Myr. The main conclusion is that star-formation occurred very recently in SECCO~1, in fact it may be currently ongoing.

 
\begin{figure}
\includegraphics[width=\columnwidth]{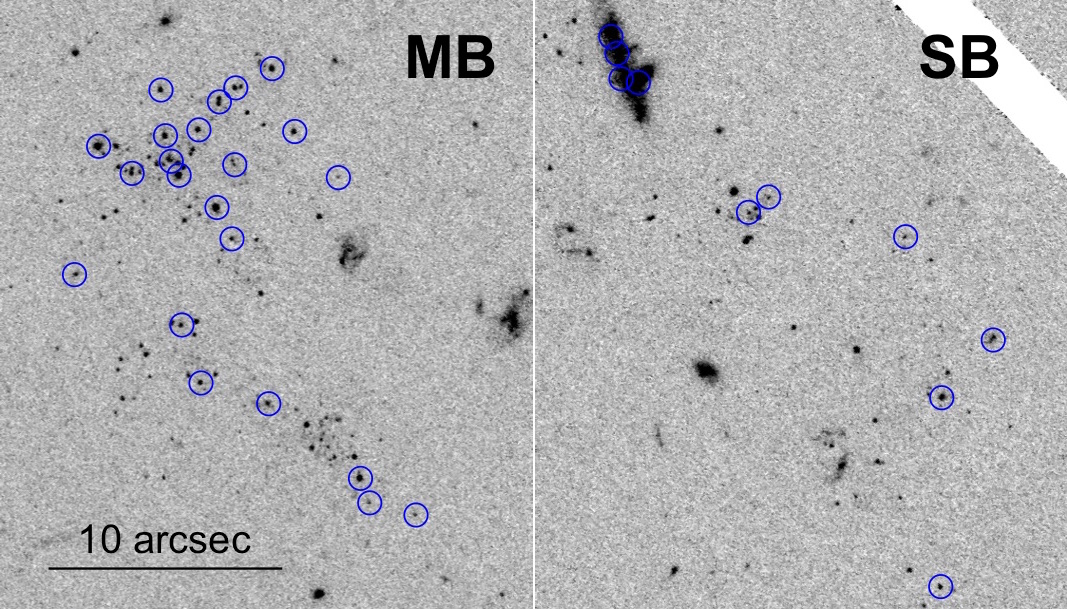}
\includegraphics[width=\columnwidth]{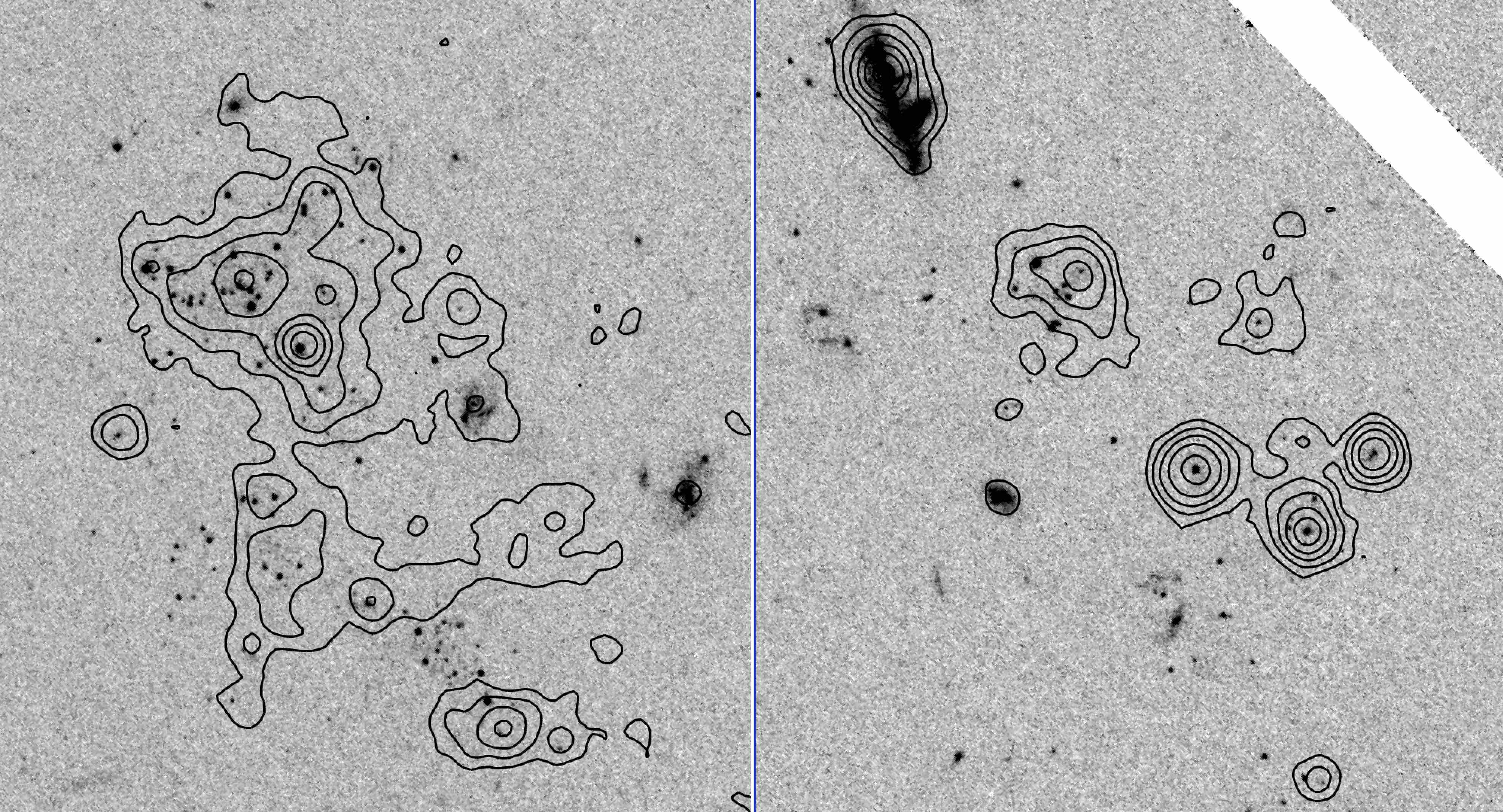}
\caption{ Upper panels: H${\alpha}$ emitting sources from Be17a having a counterpart in our catalog from HST photometry are plotted as blue circles over two portions of the F606W ACS image of SECCO~1-MB (left panel) and of SECCO~1-SB (right panel). Lower panels: H${\alpha}$ intensity contours over-plotted on the same image, with the same scale. The contour levels are, from the outermost to the innermost: 0.5, 1.0, 2.0, 4.0, 8.0, 16.0 $\times 10^{-18}$~erg~cm$^2$~s$^{-1}$. In all the panels North is up and East to the left.}
\label{hsth2}
\end{figure}

\section{HST+MUSE: the stellar content of SECCO~1} 
\label{cmd}

Using synthetic CMDs  from stellar models, S17 concluded that the CMD they obtained from HST data is compatible with continuous star formation between 50 and 7 Myr ago. They found no evidence of older stars associated to SECCO~1 in their CMD and they established that the observed diffuse light can be accounted for by the unresolved low-mass counterpart of the 7-50~Myr old population. Unfortunately, the actual upper limit that S17 were able to put on the total mass in old stars is very weak, M$_{\star}<6\times 10^5$~M$_{\sun}$, i.e., the data are still compatible with the presence of an old stellar population comparable to or even larger than the young component, in mass. 

Here we re-analyse the stellar content of SECCO~1 in the light of the results obtained with MUSE, using the independent photometry we obtained from the S17 data, as described in Sect.~\ref{datared_hst}.   
In Fig.~\ref{cocm} we show the CMD of four circular regions of the ACS-WFC field with radius=$20\arcsec$, corresponding to $\simeq 1.6$~kpc at D=17.0~Mpc. The first two regions are centred on SECCO~1 MB and SB, respectively. The third is centred on a dwarf spheroidal galaxy lying in the Virgo cluster that was serendipitously included in the ACS-WFC images of SECCO~1, Dw~J122147+132853 after S17\footnote{In the left panel of Fig.~14 by Be16 Dw~J122147+132853 can be seen $\simeq 110\arcsec$ to the West of SECCO~1 SB. It looks similar to other small LSB Virgo dwarfs identified in that paper but we do not include it in the Be16 list because it is partially superimposed on a distant disk galaxy, making its classification ambiguous.}. The fourth region samples a portion of the field far away from any resolved stellar system, as a Control Field (CF) providing an idea of the properties of the back/foreground population that should contaminate the CMDs of SECCO~1. 

The CMD of the Control Field seems populated by the typical intra cluster (IC) population of Virgo, consisting of a Red Giant Branch (RGB, stars with age $\ga 1-2$~Gyr) and a broad vertical plume of younger blue stars \citep[see, e.g.][discussion and references therein]{williams}.  We used the deep CMD of Virgo IC stars by \citet{williams} to derive the mean locus of IC blue stars (plotted here as a thick vertical segment) and the thick polygonal contour approximately enclosing the IC RGB population near the RGB Tip. Note that, according to \citet{williams}, the intra cluster RGB stars in their sample span the metallicity range $-2.3\le {\rm [M/H]}\le 0.0$. The limiting magnitude of our photometry is a strong function of color and runs nearly parallel to the right side of the polygonal contour. 
Hence the reddest stars of the IC RGB population are not seen here because they are beyond our detection limit.
This implies that a population of metal-rich (Z$\ga 0.004$) RGB stars associated with SECCO~1 may have gone undetected because the photometry is too shallow. 

On the other hand, it is easy to see that the RGB stars dominating the CMD of the Dw~J122147+132853 dSph are, in average, bluer than their IC counterparts, suggesting that the mean metallicity of the dSph is lower than that of the old IC population in this specific line of sight. Note that the clear detection of the RGB Tip of Dw~J122147+132853 around F814W$\sim 27.0$ allows us to unambiguously locate it at the distance of the Virgo cluster. Since the dSph appears to contain only old stars, its CMD can be considered as an additional Control Field for F814W$<26.5$.

   \begin{figure*}
   \includegraphics[width=\textwidth]{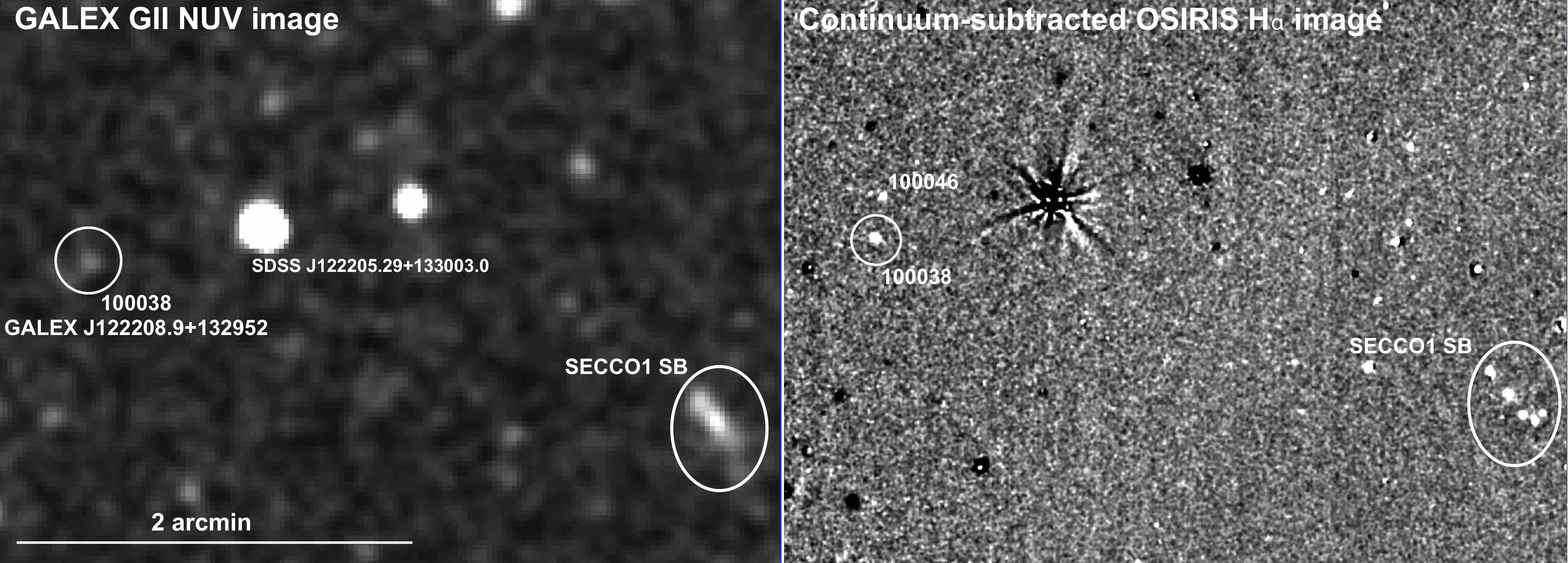}
     \caption{The newly identified candidate H${\alpha}$ emitter in the surroundings of SECCO~1. Left panel:
     GALEX NUV image (with a Gaussian smoothing of 3~px size applied) with the source (100038 / GALEX J122208.9+132962) labeled. A nearby bright star is also identified for reference, as well as SECCO~1 SB. Right panel: continuum-subtracted H${\alpha}$ image from our GTC-OSIRIS observations. In addition to 100038 and SB, here we labelled the possible companion source 100046. Note that there are sources (outside SB) that have a significant residual in the H${\alpha}$ image but have no counterpart at all in GALEX (and vice versa). In both images North is up and East to the left.} 
        \label{newsource}
    \end{figure*}

\begin{table*}
   \caption{Additional candidate H{\sc ii} regions possibly associated to SECCO~1}
\begin{tabular}{|l|l|c|c|c|c|c|c|}
\hline
  \multicolumn{1}{|c|}{ID} &
  \multicolumn{1}{c|}{GALEX ID} &
  \multicolumn{1}{c|}{RA$_{J2000}$}  &
  \multicolumn{1}{c|}{Dec$_{J2000}$} &
  \multicolumn{1}{c|}{mag$_{NUV}$} &
  \multicolumn{1}{c|}{g$^a$} &
  \multicolumn{1}{c|}{r$^a$} &
  \multicolumn{1}{c|}{H${\alpha}$ flux$^b$} \\
\hline
           &                 &  deg       &     deg     &  mag       &   mag       &     mag       & $10^{-18}$~erg~cm$^{-2}$~s$^{-1}$ \\
\hline
    100038 & J122208.9+132952 & 185.5373356 &	13.4980710 &$23.64\pm 0.42$ & $22.430\pm 0.076$ & $22.133\pm  0.041$ & $\simeq 210$  \\
    100046 &                  & 185.5365535 &	13.5015474 &	 	  & $23.543\pm 0.087$ & $22.994\pm 0.008$ & $\simeq 140$  \\
\hline
\multicolumn{8}{l}{$^a$ Sextractor Kron magnitudes ({\tt MAG\_AUTO}) from the deep images of \citet{pap1}.}\\ 
\multicolumn{8}{l}{$^b$ Assuming the same distance of SECCO~1, D=17.0~Mpc.}\\ 
\end{tabular}
\label{newcand}
\end{table*}

The CMDs of both MB and SB show an obvious excess of blue stars (F606W-F814W$<0.0$) and of red stars brighter than F814W$=25.5$ with respect to the dSph field and the CF. These stars are the characteristic population of SECCO~1, as witnessed also by the significant fraction of them being cross-identified with peaks of H${\alpha}$ emission found by B17a in the MUSE images (blue circles). The identification of sources associated with H{\sc ii} regions in the CMD of Fig.~\ref{cocm} is an important development with respect to the analysis by S17, with implications on the interpretation of the diagram (see below). The position of the MUSE sources in the ACS image is illustrated in the upper panels of Fig.~\ref{hsth2}, while in the lower panels of the same figure we over-plotted the H${\alpha}$ intensity contours.
There are several interesting considerations regarding the CMDs of Fig.~\ref{cocm}:

\begin{enumerate}
\item We confirm the conclusion by Be16, Be17a and S17 that no trace of excess RGB population  is seen around SECCO~1. Unfortunately, the GO-13735 images are so shallow that only the very tip of the RGB of metal-poor/intermediate populations at D$\simeq 17.0$~Mpc can be (barely) reached, while the tip of metal-rich populations is below the detection threshold. Deeper photometry would allow to draw much firmer conclusions on this crucial issue. However, based on lack of detection also in diffuse light (S17), in the following, as a working hypothesis, we will assume that the star formation in SECCO~1 started $\la 50$~Myr ago.

\item The apparent lack of an RGB associated with SECCO~1 prevents a robust estimate of the distance of the system, that remains quite uncertain. For example a distance as large as D=30~Mpc, as unlikely as it may appear, cannot be firmly excluded with the existing photometric data. The luminosity of the O stars associated with the H{\sc ii} regions broadly imply D$\ga 10$~Mpc (see Sect.~\ref{laura}). The strong deviation from the Hubble flow at distances larger than $\simeq 3$~Mpc \citep[see, e.g.][]{mcc} remains the most robust argument for SECCO~1 to belong to the Virgo cluster (see B15b) and to the LVC, in particular (S15, B17a, S17). 

\item We tentatively fit the CMD of SECCO~1 with two isochrones from the most recent PARSEC set \citep{parsec}, with the mean metallicity  of SECCO~1, as measured by Be17a, assuming D$=17.0$~Mpc. The isochrones of age 4~Myr (light grey) and 35~Myr (dark grey) appear to nicely bracket the stellar population that can be unambiguously associated to SECCO~1.  The comparison with the dSph and CF CMDs suggests that there is no compelling evidence for a significant population older than $\simeq$35~Myr (but see S17). We confirm that there is no obvious difference in stellar population between MB and SB (S17).

\item \citet{reines} showed that the spectrum of the ionised gas can strongly affect the broadband photometry of massive stars within H{\sc ii} regions. Both the continuum and strong emission lines can have a sizeable impact (note, e.g., that the F606W passband fully encloses both the [OIII] and H${\alpha}$ lines, that are seen to vary by a considerable amount from region to region in SECCO~1, see Be17a). Hence interpreting the position in the CMD of stars associated with H{\sc ii} regions (large blue circles in Fig.~\ref{cocm}) with stellar isochrones may lead to misleading conclusions. 
This kind of effect is likely at the origin of the fact that some stars associated with H{\sc ii} regions have colors $F606W-F814W\sim 1.0$, typical of stars with $T_{eff}\sim 4000$~K, while the observed nebular spectra require a ionising source of much higher temperature ($T_{eff}\ga 30000$~K, see Sect.~\ref{laura}).   
Since some contribution from nebular spectra may affect also the stars not directly associated with a H${\alpha}$ peak detected in the MUSE data-cubes (due, e.g., to diffuse emission, see Fig.~\ref{hsth2}), the overall CMD of SECCO~1 must be interpreted with some caution.  

\item We have marked with a skeletal triangle symbol the MUSE sources that are recognised by DOLPHOT as significantly more extended than point sources of the same magnitude, according to the sharpness parameter. Most of them are fainter than F814W=26.5 while most of the brightest MUSE sources appear as normal stars. This implies that, in general, sources associated with H{\sc ii} regions of SECCO~1 are unresolved in the available ACS images (i.e. are point-like) and blending is the most likely reason why many of the faintest sources appear slightly extended. 
\end{enumerate}

\section{H$\alpha$ imaging: searching for additional pieces}
\label{pieces}

It is likely that some of the candidate H$\alpha$-emitting sources selected in Sect.~\ref{datared_gtc} from the GTC images are not related to SECCO~1. Background galaxies that happen to have strong emission in the wavelength window of the f657/35 filter are an obvious example.
Since SECCO~1 sources are clearly seen also in GALEX NUV images from the GII survey, we decided to validate our candidates by asking that they are detected also in these images. 
Indeed we found a source with strong H${\alpha}$ excess that is detected in the GALEX NUV image enclosing SECCO~1, $\simeq 3.3\arcmin$ to the East-North-East of SB, as shown in Fig.~\ref{newsource}. 
The source (ID number 100038 in our catalog) is seen as extended both in the OSIRIS images and in the deep g,r images of the original SECCO survey \citep{pap1}. It has a nearby point-like companion (100046) with significant H${\alpha}$ excess. Despite the lack of a GALEX counterpart we include also this source in the list of candidate additional H{\sc ii} regions possibly associated with SECCO~1 (Tab.~\ref{newcand}) because of its extreme proximity to 100038 ($12.7\arcsec$).

We derived a rough estimate of the H${\alpha}$ flux of 100038 and 100046 by comparison with SECCO~1~SB sources that have been measured in Be17a, assuming that the newly found candidates lie at the same distance. The derived fluxes are similar to those of confirmed SECCO~1 sources (Be17a). It is important to note that 100038 and 100046 lie just beyond the circle of radius $4.0\arcmin$ centred on the centre of the OSIRIS field of view that is safe from any contamination of other interference orders and should be considered as the actual operative field of view of OSIRIS tunable filters. Moreover, all the known SECCO~1 sources that are seen in the OSIRIS images are not detected or very weak in the continuum images, while the continuum flux of the two candidates identified here is clearly detected. Still, especially given the GALEX detection of 100038, they remain good candidate H{\sc ii} regions. 

While a spectroscopic follow-up is needed to ascertain the nature of 100038 and 100046, as well as other 
H${\alpha}$ sources lacking a GALEX counterpart, the overall result of our search for additional star forming regions associated to SECCO~1 is largely negative. In particular, we did not found any reliable candidate in the region suggested by Be16 as possibly hosting additional SECCO~1 sources (see their Fig.~14; the region is partially vignetted by the gap between the two chips of OSIRIS). The operative OSIRIS field samples a projected circular area with radius $\simeq 20$~kpc around SECCO~1 at the distance of D=17~Mpc, still only one candidate 
with properties analogous to known SECCO~1 pieces has been found.

\section{The nature of SECCO~1}
\label{disc}

\begin{figure}
   \includegraphics[width=\columnwidth]{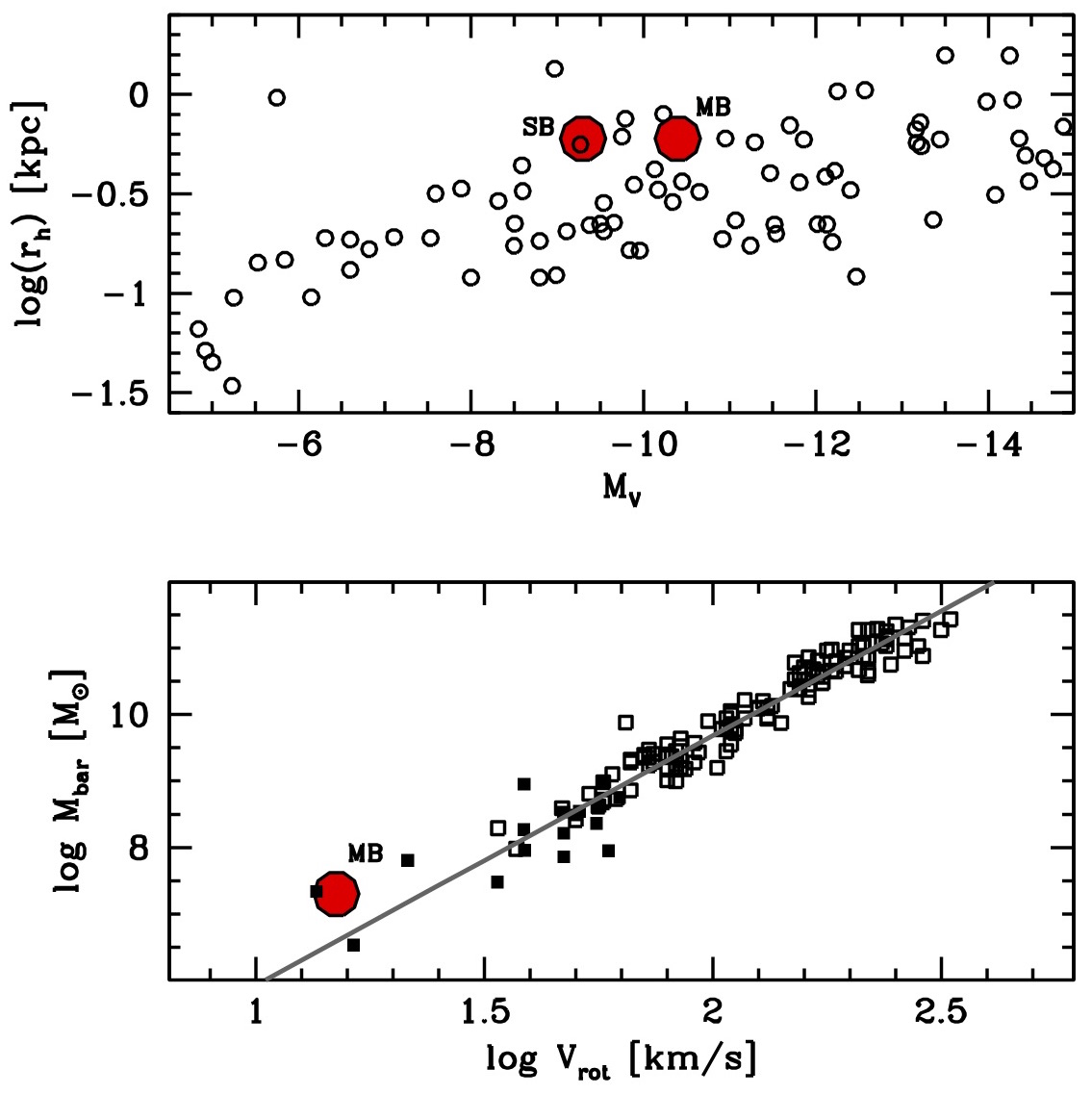}
     \caption{Upper panel: SECCO~1 MB and SB (large red filled circles) in the integrated absolute V magnitude vs. log of the half-light radius diagram. Small empty circles are dwarf galaxies in the Local Group from \citet[][]{mcc}. As a proxy for the half-light radius for MB and SB we adopted half of the radial size reported (and defined) in Tab.~\ref{mean}.
     Lower panel: baryonic Tully-Fisher relation. Open squares are from \citet{lelli16} and filled squares from \citet{iorio17}; the grey line is the best-fit for the accurate distances sample by \citet{lelli16}. The baryonic mass of SECCO~1 MB (large red filled circle) has been computed by summing the H{\sc i} mass corrected for the contribution of helium to the total gas mass according to \citet{lelli16}; the rotation velocity is from A15 that provide an estimate of this quantity only for MB. } 
        \label{tfrh}
    \end{figure}

In Fig.~\ref{tfrh} we show SECCO~1 MB and SB in the $M_V$ vs. log(r$_h$) relation for dwarf galaxies in the Local Volume \citep[from][]{mcc} and in the baryonic Tully-Fisher relation \citep[see][and references therein]{tay13,lelli16,iorio17}.
Only MB is included in the last diagram since only for this component A15 provide an estimate of $V_{rot}$ based on the velocity gradient seen in H{\sc i} (see also Be17a). The two diagrams show that, taking the observables at face value, the structural and dynamical properties of SECCO~1 are in the range of those expected for a dwarf galaxy of the same baryonic mass. 

On the other hand, the velocity field of the overall system (gas + stars), and the off-set in position between SB and AGC~229490 suggest that it is unlikely that the system  is in dynamical equilibrium (see also Sect.~\ref{simu}, below). If confirmed, also the lack of an underlying old population would argue against the hypothesis that SECCO~1 is an ordinary, albeit very dark, dwarf galaxy \citep[see][for a theoretical framework for the evolution of almost-dark galaxies similar to SECCO~1]{bekki15}.

   \begin{figure}
   \includegraphics[width=\columnwidth]{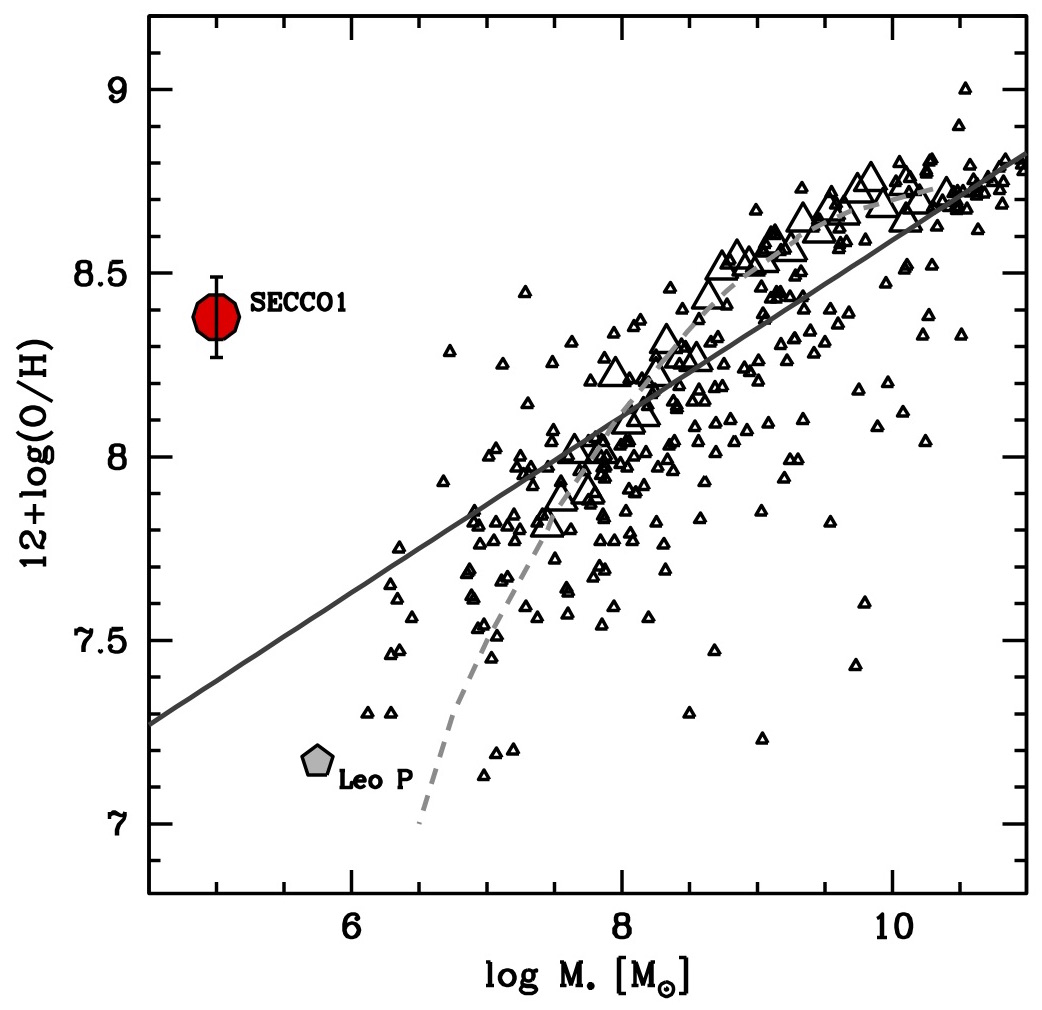}
     \caption{SECCO~1 as a whole into the stellar mass - metallicity relation. Small triangles are the z$\simeq 0$ sample of \citet{hunt16}, large triangles are from \citet{am13}, the light grey long-dashed curve is the best-fit to the same data (from direct-method oxigen abundance estimates) extrapolated to low masses, from the same source. The dark grey continuous line is the stellar mass - metallicity relation from planetary nebulae by \citet{gon14}. SECCO~1 is represented as a large red filled circle.} 
        \label{massmet}
    \end{figure}

However, the strongest, and likely ultimate, argument against this hypothesis is provided by Fig~\ref{massmet}, which displays how far SECCO~1 is from the stellar mass - metallicity relation of normal galaxies (see B15a, Be17a, S17). It is clear that such a high (and extremely homogeneous) metallicity cannot arise from chemical evolution within a stellar system of such low mass. The stripping of some pre-enriched gas from, e.g., the disc of a spiral galaxy appears as the most likely origin for the system, as it would fit both the high metallicity and  the (possible) lack of an old stellar population. There is no obvious way of distinguishing between a tidal stripping or a ram pressure stripping event for the origin of SECCO~1. Indeed, as noted by B17a, the location of SECCO~1 in Fig.~\ref{massmet} is typical of tidal galaxies \citep[see, e.g.][and references therein]{sweet,Duc}. However, star-forming knots similar to SECCO~1 are sometime seen to originate in ram pressure stripping episodes within galaxy clusters \citep[{\em fireballs}, see, e.g.,][references and discussion therein]{yoshi,fuma,kenney} and are predicted to occur in hydrodynamical simulations \citep{kap09}. 

\subsection{Comparison with fireballs}
\label{fireb}

\citet{yoshi} and \citet{hester,fuma,kenney} provide very detailed analyses of star-forming knots within H${\alpha}$ tails likely arising from strong ram-pressure stripping events in galaxy clusters.
All these authors refers to these compact UV and H${\alpha}$ emitting blobs as {\em fireballs}. The similarities of SECCO~1 with fireballs was already briefly noted in B15a; here we make a more thorough comparison.  

\citet{yoshi} show that fireballs have emission line-ratios typical of H{\sc ii} regions and the same metallicity as the gas in the parent galaxy. They have typical sizes similar to SECCO~1 (1-2~kpc, but with $r_h\simeq 200-300$~pc) and similar or larger stellar masses \citep[e.g., those considered by][have $3.9\times10^{4}~M_{\sun}\le M_{\star}\le 5.0\times 10^{5}~M_{\sun}$]{fuma}. Typical mean ages seem larger than SECCO~1, in the range $\sim 80-400$~Myr, but (g-r) colors are similar to SECCO~1 and the presence of H{\sc ii} regions imply very recent / ongoing star formation. \citet{fuma} report a star formation rate of $10^{-3} - 10^{-4}~M_{\sun}{\rm yr}^{-1}$, again pretty similar to SECCO~1. \citet{kenney} report on the detection of a single unresolved H{\sc i} cloud with mass $3.7\times10^7~M_{\sun}$, not clearly associated with an individual fireball but having systemic velocity similar to several H{\sc ii} regions in the ram-pressure tail of IC~3418. 

While many intrinsic properties of fireballs are quite similar to those of SECCO~1, there is also at least one remarkable differences. Fireballs are found within a few tens of kpc from their parent galaxy, within obvious cometary tail structures, and they ignited the star-formation as soon as they were stripped.  
On the other hand the cloud that formed SECCO~1 traveled for a long time away from its parent galaxy and it is currently forming star in conditions of extreme isolation.

   \begin{figure}
   \includegraphics[width=\columnwidth]{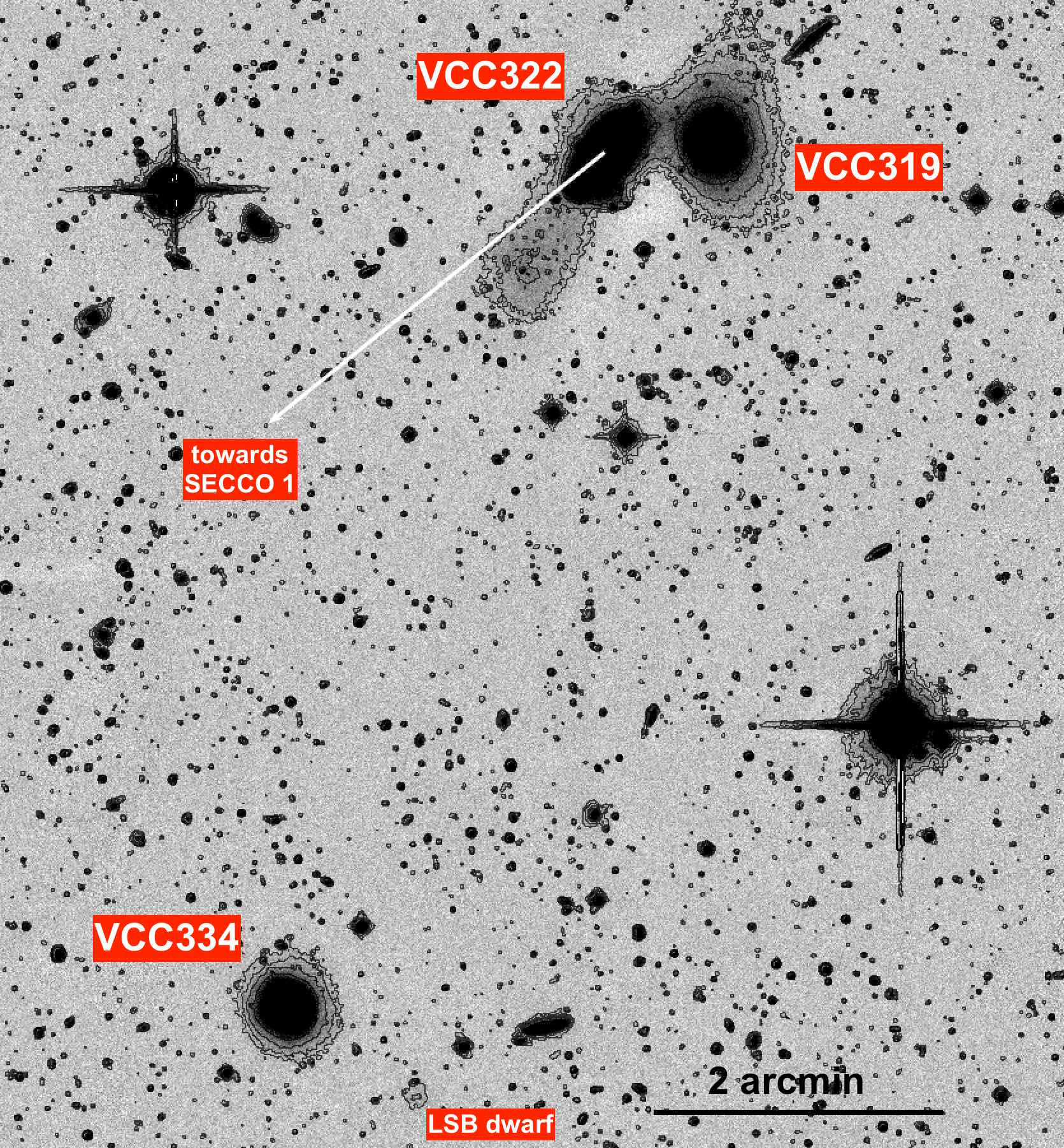}
     \caption{Deep g-band image of the interacting group of dwarfs suggested as candidate site of origin of SECCO~1, from the Next Generation Virgo cluster Survey \citep{ferra}. North is up and East is to the left. Density contours at arbitrary intensity levels have been drawn to put in evidence low surface brightness substructures, like the prominent tidal tail of VCC~322 or the very low SB dwarf to the West-South-West of VCC~334 (labelled).  The white arrow shows the direction from VCC~322 towards SECCO~1 MB.} 
         \label{ami}
   \end{figure}

\subsection{Candidate parent galaxies}
\label{candi}

B17a proposed the interacting galaxy pair formed by NGC~4299 ($V_h=+258\kms$) and  NGC~4294 ($V_h=+369\kms$) \citep{chung} as a possible site of origin for the gas cloud that is now SECCO~1
\footnote{All the parameters of candidate parent galaxies reported in this section are taken from the SIMBAD database {\tt http://simbad.u-strasbg.fr}.}. The gas distribution of these galaxies shows signs of the effects of both ram-pressure and tides and their velocities are compatible with the range of LVC galaxies, but they are more than 500~kpc apart from SECCO~1, in projection (see Tab.~\ref{mean}). S17 noted that the M~86 ($V_h=-183\kms$) group is also member of the Virgo LVC and it is located at $\simeq 350$~kpc, in projection, from SECCO~1. Galaxy-galaxy interactions are clearly ongoing and, in particular, M~86 is likely interacting with NGC~4438 ($V_h=+98\kms$) and both seems to be undergoing ram-pressure stripping, with a short X-ray and H${\alpha}$ emitting tail generically pointing in the direction of SECCO~1. We note that in the surroundings of the same group an extended H{\sc i} tail is probably being stripped from NGC~4388 ($V_h=+2555\kms$) by its interaction with M~86 in a direction nearly perpendicular to that toward SECCO~1 \citep{tom}, showing that this environment is particularly active as a source of stripped H{\sc i} clouds.

In Fig.~\ref{ami} we present a new promising candidate site for the origin of SECCO~1, the group of dwarf galaxies VCC~322 ($V_h=-438\kms$), VCC~319 ($V_h=-209\kms$) and VCC~334 ($V_h=-213\kms$)\footnote{The NGSV image shown in  Fig.~\ref{ami} reveal the presence of a very low surface brightness amorphous galaxy located $\simeq 70\arcsec$ to the West-South-West of VCC~334. This may be an additional member of the group.}. These have velocities typical of LVC members and are likely mutually bound and interacting. Two of them are gas-rich and star-forming (VCC~322 and VCC~334) while the other is red and with no H{\sc i} detected. Their absolute integrated r magnitudes are all enclosed between $M_r=-15.9$ and -16.1. The stellar masses, derived as in \citet{micudg} adopting the color vs. mass-to-light relations by \citet{rc15}, are 
$M_{\star}\simeq 9.0\times 10^{7}~M_{\sun}$, $3.8\times 10^{8}~M_{\sun}$, and $8.3\times 10^{7}~M_{\sun}$ for VCC~322, VCC~319, and VCC~334, respectively.
In projected distance they are closer to SECCO~1 than the candidates described above ($\simeq 250$~kpc, see Tab.~\ref{mean}). 

However,  the feature that makes them particularly promising as candidate parents for SECCO~1 is their chemical composition. Both VCC~322 and VCC~334 have good spectra in the range 3800\AA$\la \lambda \la$ 9200\AA~ from the Sloan Digital Sky Survey - Data Release 12 \citep{dr12} with measured fluxes for several strong emission lines\footnote{See {\tiny\tt dr12.sdss.org/spectrumDetail?plateid=1766\&mjd=53468\&fiber=56} and {\tiny\tt
dr12.sdss.org/spectrumDetail?plateid=1766\&mjd=53468\&fiber=50} for VCC~322 and VCC~334, respectively.}.
We estimated the oxygen abundance of these two galaxies in the same way as we did for SECCO~1 in Be17a, averaging the results of the indicators N2 and O3N2 from \citet{pp04}. For both galaxies we obtain ${\rm 12+log(O/H)}=8.3\pm 0.2$, i.e. the same abundance of SECCO~1, within the uncertainties (that here include observational errors and uncertainties in the calibrations of the relations between oxygen abundances and strong-line indicators), and consistent with the range of abundances of galaxies with that stellar mass, albeit slightly on the high side, possibly suggesting that they indeed lost some mass in the past. 
Adopting the linear stellar mass - metallicity relation by \citet{gon14}, shown in Fig.~\ref{massmet}, ${\rm 12+log(O/H)}=8.10$, $8.09$, and $8.25$ is obtained for VCC~322, VCC~334, and VCC~319, respectively. 

A close two/three body encounter may have produced the required stripping episode. It is tantalising to imagine that the same episode transformed VCC~319 into the passive gas-less system that we see today. 

\section{Hydrodynamical simulations of a pressure-supported cloud}
\label{simu}

B17a estimated that if the gas cloud that gave rise to SECCO~1 detached from the interacting galaxy pair 
NGC~4299 + NGC~4294, it should have travelled for at least $\sim 1.2$~Gyr within the LVC ICM to reach its current position, given the projected distance to the pair and the radial velocity difference. Analogous computations imply travel times $>1.0$~Gyr also from the other proposed parent galaxies or galaxy groups.
Hence, the extreme isolation of SECCO~1 implies that the system has survived within its environment for a long time before the onset of star formation. Is this occurrence realistic? Are such kind of clouds expected to have lifetimes longer than 1~Gyr? To attempt an answer to the last question we performed a two-dimensional high-resolution hydrodynamical simulation of a cold gas cloud travelling through a hot and low-density ambient medium, with the  typical features of the LVC ICM. 

\subsection{Simulation set up}
\label{Setup}

The simulation was carried out with the ATHENA code \citep{stone08}, using a two-dimensional Cartesian geometry and a fixed grid with size 40 kpc~$\times$~40 kpc and spatial resolution 5 pc~$\times$~5 pc. We included a module for the classical thermal conduction \citep{spitzer} with an efficiency of $10\%$ (accounting for the presence of tangled magnetic fields) and a module for radiative gas cooling and heating in the presence of collisional ionisation, and photoionisation from an uniform UV background at redshift $z=0$. Cooling and heating rates have been obtained through the CLOUDY spectral synthesis code \citep{ferland13} and the spectrum of the background radiation field has been taken from \citet{haardt}. The temperature cutoff for radiative cooling is $10^3$~K, then molecular cooling is not included in our simulation. For further details on the hydrodynamical treatment of thermal physics we refer to \citet{armillotta16, armillotta17}. 
Moreover, we used an adaptive and moving grid that follows the cloud during its own motion. We neglected the presence of an external gravitational potential and self-gravity inside the cloud (we justify the latter assumption in Sec.~\ref{Results}).

\begin{table*}
  \begin{center}
  \caption{Initial parameters of the simulation: hot gas temperature $T_{hot}$, hot gas density $n_{hot}$, hot gas metallicity $Z_{hot}$, cloud temperature $T_{cl}$, cloud density $n_{cl}$, cloud metallicity $Z_{cl}$, cloud radius $R_{cl}$, cloud velocity $v_{cl}$.}
  \label{ParSim}
  \begin{tabular}{ccccccccc}
\hline
$T_{hot}$& $n_{hot}$& $Z_{hot}$& $T_{cl}$& $n_{cl}$ & $Z_{cl}$&$R_{cl}$&$M_{cl}$ & $v_{cl}$\\
(K) & (cm$^{-3}$) & (Z$_{\odot}$) &(K) & (cm$^{-3}$) &  (Z$_{\odot}$)&(kpc)&($\mo$)&($\kms$)\\
\hline
$5\times10^6$& $2.5\times10^{-5}$&$0.1$ &$5\times10^3$&$2.6\times10^{-2}$&$0.5$&$3.7$ &$9.4\times10^7$&$200$\\
\hline	      
\end{tabular} 
\end{center}
\end{table*}

\begin{figure*}
\includegraphics[width=\textwidth]{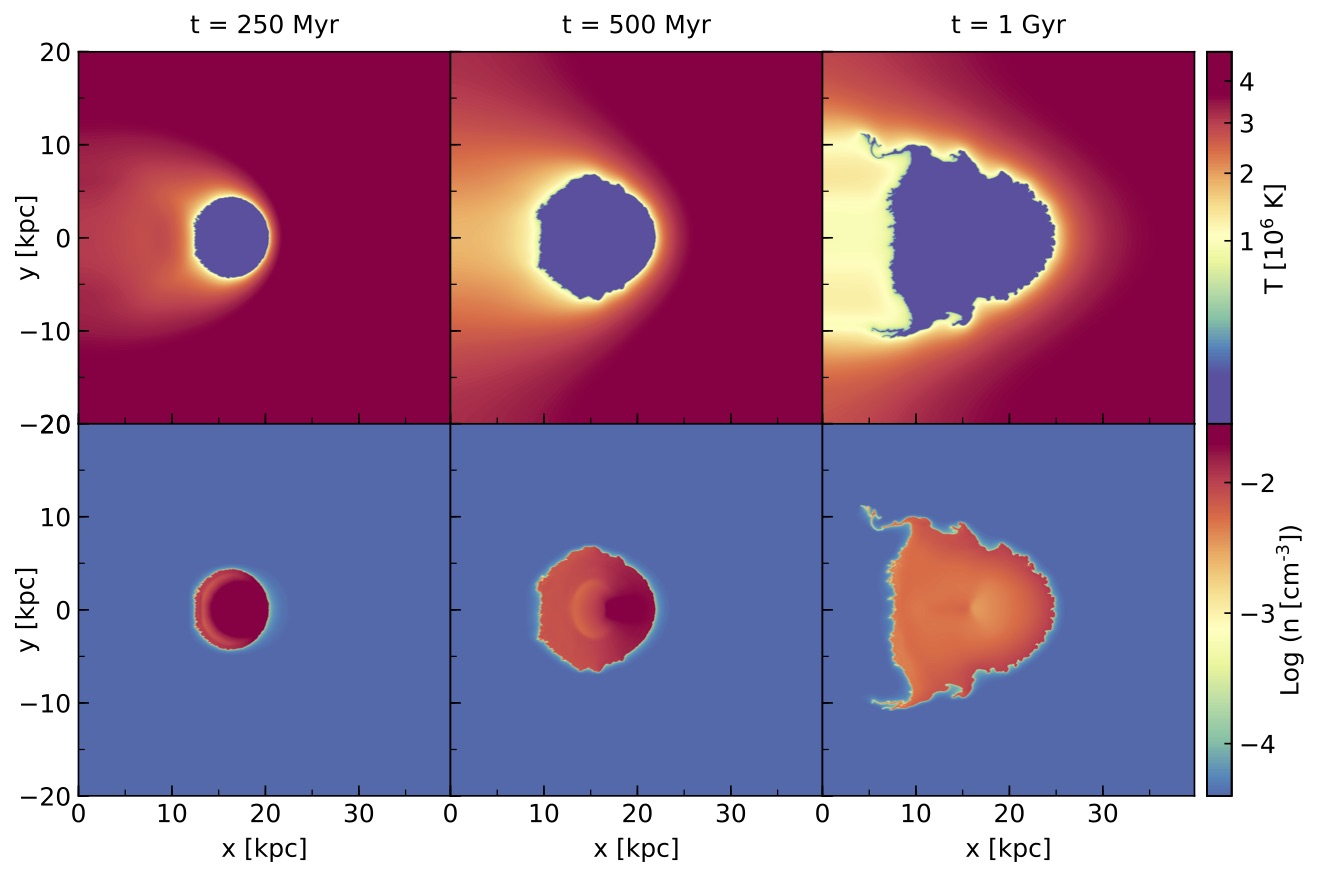}
\caption{Temperature (top panels) and number density (bottom panels) snapshots of the simulation at t=250 Myr (left panels), t=500 Myr (middle panels) and t=1 Gyr (right panels) for the 2D simulation.}
\label{Snapshot}
\end{figure*}

The parameters of the simulation are listed in Table~\ref{ParSim}. The temperature of the hot gas, $T_{hot} = 5 \times 10^6$~K, was calculated through the following formula:

\begin{equation}
T_{hot} = \dfrac{3 \sigma_{r}^2 \mu_{hot} m_{p}}{2 k_{B}}\;,
\end{equation}

\noindent
where $\mu_{hot} \simeq 0.59$ is the mean molecular weight of the hot gas, $m_{p}$ the proton mass, $k_{B}$ the Boltzmann constant and $\sigma_{r} \sim 208~\kms$ is the radial velocity dispersion of the ICM in LVC \citep{boselli}. The hot gas metallicity was set to $0.1$~Z$_{\odot}$, according to the metallicity of the ICM at 1.1~Mpc from the center of the Virgo cluster \citep{urban}, where 1.1 Mpc is the projection on the sky of the physical distance between NGC4216, the central galaxy of LVC, and M87, the central galaxy of Cluster A.
Finally, the numerical density of hot gas was set to $2.5\times10^{-5}$~cm$^{-3}$. This value was derived from the total mass (dark and baryonic matter) density in LVC. For this calculation, we assumed that the hot gas density distribution follows the dark matter density profile given by \citet{navarro}:  

\begin{equation}
{\rho (r)}  =  {\rho_{cr}} \,\dfrac{\Delta_{vir}} {3} \dfrac{c_{vir}}{f(c_{vir})}\dfrac{r_{s}^3}{r (r + r_{s})^2} \; ,
\label{DM}
\end{equation}

\noindent
where $ {\rho_{cr}}$ is the critical density of the Universe and $\Delta_{vir}$ is the virial overdensity provided by the dissipationless spherical top-hat collapse. In the $\Lambda$CDM cosmological model with $\Omega_{m}=0.3$ the value of $\Delta_{vir}$ is $\sim 340$ at $z=0$ \citep{bryan}.
Moreover, $c_{vir}$ is the concentration parameter associated to the mass, $M_{vir}$, contained within the virial radius, $r_{vir}$, $r_{s}\equiv r_{vir}/c_{vir}$ is a scale radius and $f(c_{vir})$ a function of $c_{vir}$ \citep[see][for definitions of $c_{vir}$ and $f(c_{vir})$ at redshift $z=0$]{duffy}. The virial mass of LVC, $M_{vir} \sim 1.2 \times 10^{13}~\mo$, has been estimated through the virial theorem:

\begin{equation}
M_{vir} =  \dfrac{3 r_{vir} \sigma_{r}^2}{G}
\label{Mvir}
\end{equation}

\noindent
where $G$ is the gravitational constant and the virial radius, $r_{vir}$, is defined as the radius of a sphere within which the average mass density is $\Delta_{vir} \rho_{cr}$. By using Eq.~\ref{DM}, we calculated the total mass density, $\rho(r)$, at 430 kpc from the center of LVC, where 430 kpc is the projection on the sky of the physical distance of SECCO~1 from the central galaxy of LVC. The mass density of hot gas was estimated by multiplying $\rho(r = 430~\mathrm{kpc})$ by the universal baryon fraction \citep[$\sim 0.17$, e.g.,][]{komatsu} by the expected fraction of hot gas \citep[$\sim 0.3$,][]{Dai}.  

For the cloud temperature we adopted a value of 5000~K, according to the mean temperature of the diffuse H{\sc i} \citep[e.g.,][]{wolfire, roy}. The cloud metallicity was set to $0.5$~Z$_{\odot}$, following B17a. The cloud velocity with respect to the surrounding medium is set at $200~\kms$, similar to the velocity dispersion of the LVC \citep[][]{boselli}. The cloud radius is 3.7~kpc (A15). We note that the cloud is fully resolved since its radius is almost three orders of magnitude larger than the resolution of our spatial grid. From the radius and the total H{\sc i} mass, we derived the H{\sc i} numerical density of the cloud, $n_{cl, HI} \simeq 2.9 \times 10^{-3}$~cm$^{-3}$. To calculate the total numerical density, $n_{cl}$, we divided $n_{cl, HI}$ by $\mu_{cl} \simeq 0.71$, the mean molecular weight of the cloud, and  $X_{HI}/A \simeq 0.16$, the mass abundance of H{\sc i} with respect to all elements at $T=5000$~K and $Z=0.5$~Z$_{\odot}$. Both these values have been obtained through our CLOUDY modelling. We found $n_{cl}\simeq2.6\times10^{-2}$ cm$^{-3}$. We note that this value of density almost corresponds to the density required for the cloud to be in pressure equilibrium with the ambient medium. Given this density and including the contribution of neutral and ionized hydrogen, helium and metals, the total mass of the cloud is $M_{cl}=9.4\times 10^7~$M$_{\sun}$. The simulation ran for 1 Gyr.

\subsection{Results}
\label{Results}

\begin{figure}
\includegraphics[width=0.5\textwidth]{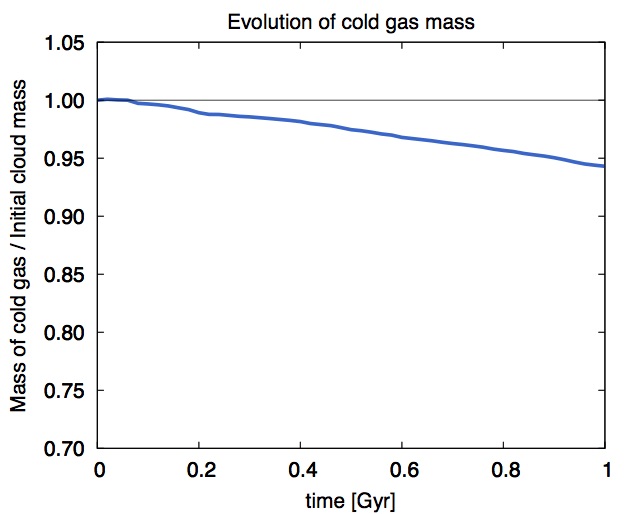}
\caption{Evolution of the cold gas ($T <2 \times 10^{4}$ K) fraction as a function of time in the 2D simulation.}
\label{Mass}
\end{figure}

Figures ~\ref{Snapshot} and ~\ref{Mass} display the results of our simulation. 
Fig.~\ref{Snapshot} shows the temperature (top panel) and number density (bottom panel) distributions on the grid at different simulation times. The cloud evolution proceeds quite slowly during 1 Gyr. Until 250 Myr, the initial spherical cloud is completely intact, and all its features (temperature, density and radius) remain nearly constant. The only difference is the presence of a parabolic layer around the cloud, characterized by temperatures that slowly increase from the cloud to the ambient temperature. This effect is explained by the presence of thermal conduction, which smooths the temperature gradients at the interface between cold and hot gas, leading the gas cloud to evaporate. After 250 Myr, hydrodynamical instabilities start to develop at the edge of the cloud, slowly causing its deformation. Moreover, the cloud expands - with an average radius that increases by more than a factor two after 1 Gyr - decreasing its density but keeping its temperature nearly constant. Indeed, in order to preserve pressure equilibrium with the ambient medium, the cloud continuously adapts to the surrounding thermodynamical changes. The gas temperature around the cloud decreases both because thermal conduction makes the temperature gradient more and more smooth, and because the gas lost from the cloud quickly evaporates creating a wake at intermediate temperature. As a consequence, the cloud density decreases, while its temperature remains constant due to thermal effects, as we discuss later.

Fig.~\ref{Mass} shows the time evolution of the mass of cold gas at $T < 2 \times 10^{4}$~K.
 Above such a temperature, the hydrogen is generally completely ionised.
The loss of cold gas mass from the cloud is negligible, it is $\sim 6\%$ of the initial
mass of the cloud after 1 Gyr.

The cloud survival strongly depends on the efficiency of ram pressure exerted by the ambient medium. External ram pressure slows down the cloud during its motion, triggering the formation of hydrodynamical instabilities and the subsequent cloud destruction. However, the efficiency of ram pressure decreases with increasing the cloud mass. The time scale associated to this phenomenon is the so-called drag time \citep[e.g.,][]{frat06}:

\begin{equation}
t_{drag}=\dfrac{M_{cl}}{v_{cl}\sigma\rho_{hot}}\; ,
\label{tdrag}
\end{equation}

\noindent
where $\sigma \simeq \pi {R_{cl}}^2$ is the cross section of the cloud. In our case, $t_{drag} \sim 28$~Gyr, more than two orders of magnitude larger than the simulation time. The relatively large mass of the cloud and the low density of the surrounding medium make the ram pressure effect very weak and justify its solid body-like behaviour.  The low density of the medium is justified by the fact that the cloud is located at the edges of the LVC halo. Indeed, the distance of the cloud  from the center of LVC, 430~kpc, is slightly larger than the virial radius, $\sim 400$~kpc (see Eq. \ref{Mvir}). 

The cloud is also stabilised by thermal effects. According to our CLOUDY cooling function, $T\sim6000$~K is the equilibrium temperature between radiative cooling and heating at numerical density $n \sim n_{cl}$. Moreover, the cloud evaporation driven by thermal conduction is very slow. According to the classical theory, the evaporation time of our cloud is $\sim 24$~Gyr \citep[][where we considered an efficiency of thermal conduction of $10 \%$]{cowie}, once again two orders of magnitude larger than the simulation time.

In order to evaluate if the presence of self-gravity could affect our results, we calculated the virial ratio of the cloud. The virial ratio is a dimensionless value that measures the gravitational binding of a cloud: clouds with virial ratio lower than 2 are gravitationally bound. For our cloud, the virial ratio is $\sim 4$, {\em indicating that the cloud is pressure-supported and globally stable against the collapse}. The main effect that the presence of self-gravity could have on the global evolution of the cloud should be to further stabilize it against the formation of hydrodynamical instabilities \citep{murray93}.  

Given all the above considerations, we can conclude that gas-clouds similar to SECCO~1 can be long-lived in the LVC environment and can travel nearly intact for hundreds of kpc from the site of their formation \citet[see also][]{burk}. The results of the N-body/hydrodynamical simulations by \citet[][discussed in more detail below]{kap09} also suggest that cold gas clouds can be long lived in the cluster medium, as ram-pressure stripped gas clumps are still alive at the end of their simulation, 0.5~Gyr after the stripping event. 
It is also interesting to note that the friction of the ICM has only minor effects. The systemic
velocity of the cloud decreases by just $\simeq 8.0\kms$ in the 1~Gyr of our simulation.

\begin{figure*}
    \centering
    \includegraphics[width=\textwidth]{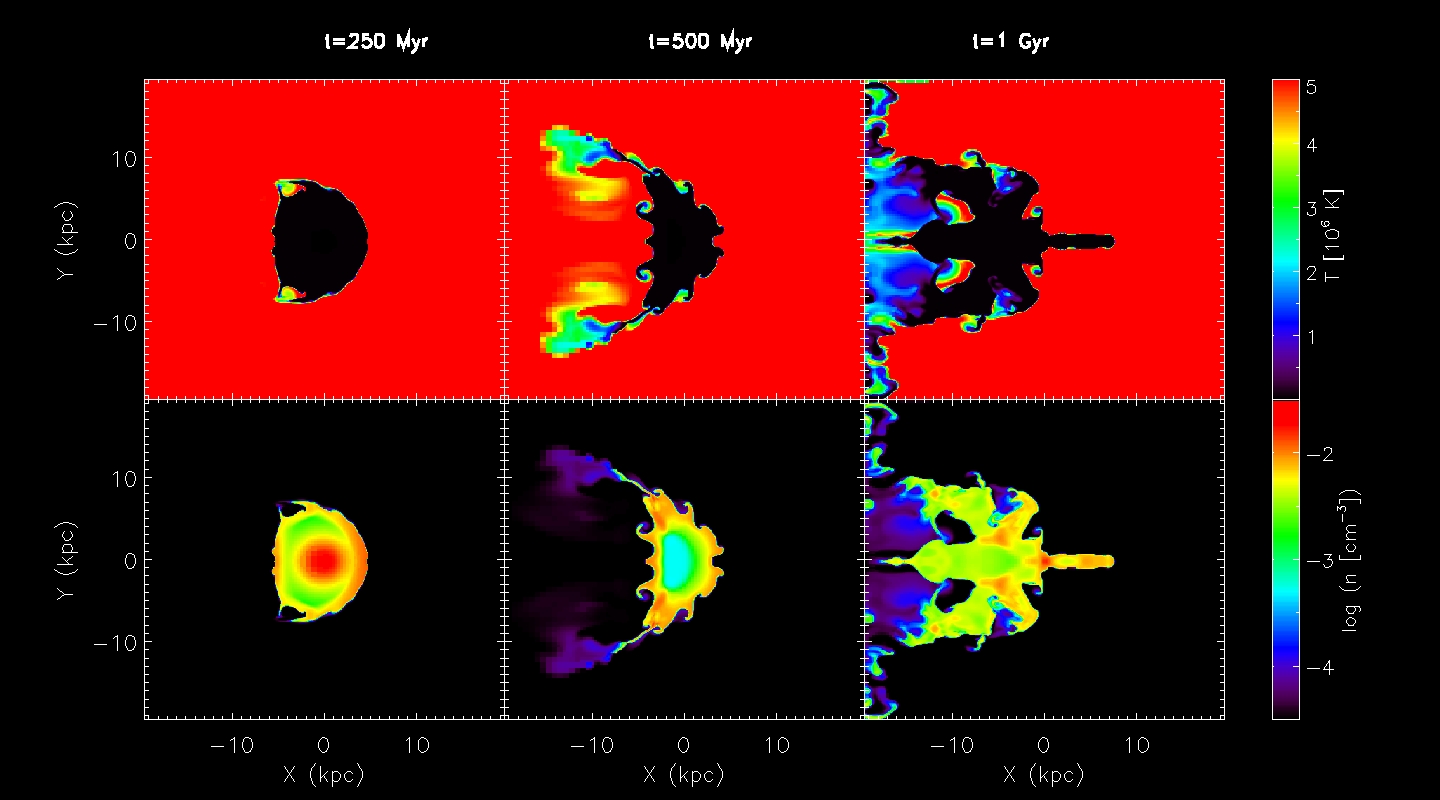}
      \caption{Temperature (top panels) and number density (bottom panels) snapshots of the simulation at t=250 Myr (left panels), t=500 Myr (middle panels) and t=1 Gyr (right panels) for the 3-D simulation.}
        \label{3dmaps}
\end{figure*}
\begin{figure}
    \centering
      \includegraphics[width=\columnwidth]{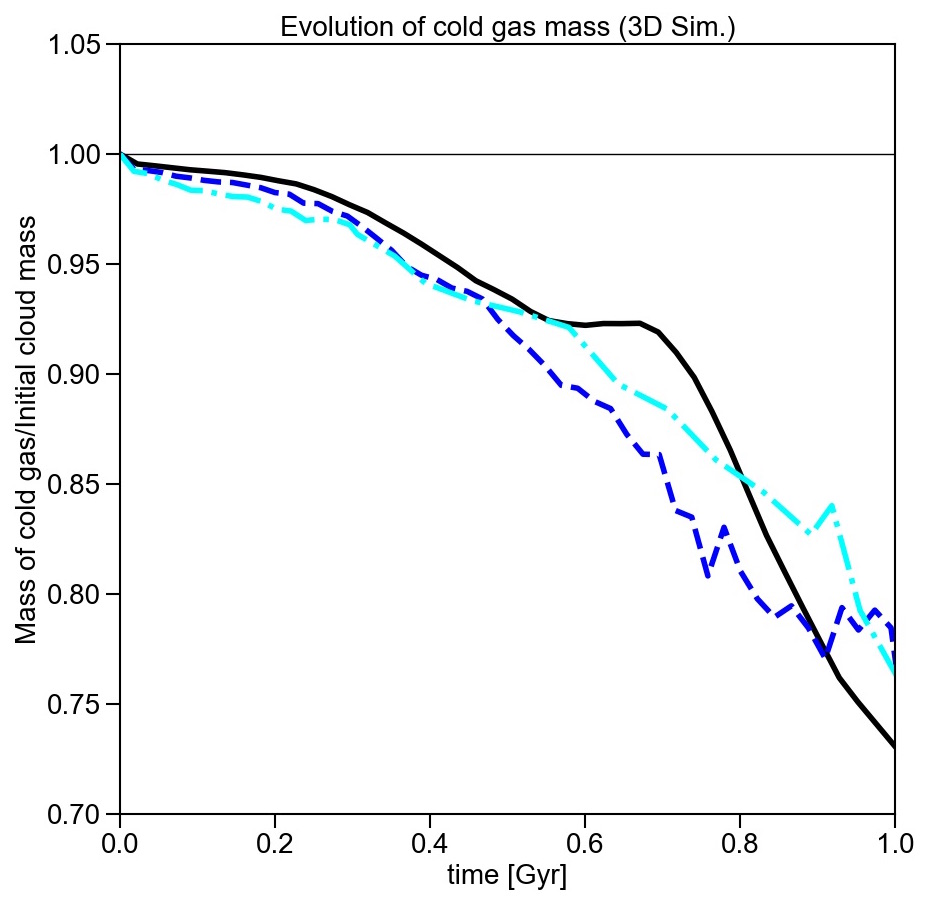}
      \caption{Evolution of the cold gas ($T <2 \times 10^{4}$ K) fraction as a function of time in the 3D simulation. The results from simulations with three different values for the
       maximum resolution are shown, as a convergency test. Dotted-dashed cyan line: $levelmax=7$, dashed blue line: $levelmax=8$, black contimuous line: $levelmax=9$. The corresponding maximum resolution is 312~pc, 156~pc, and 78~pc, respectively.}\label{cold}
\end{figure}

\subsection{A lower resolution simulation in three dimensions}

As an independent validation of the results of our 2D simulation, we ran a lower resolution, three dimensional hydrodynamic
simulation of a cold gas cloud flowing in a hot medium, adopting a very similar set-up. 
We used  a customised version of the grid-based, Adaptive Mesh Refinement (AMR) hydro-code RAMSES \citep{tes02}.
In our three-dimensional Cartesian Grid, the computational box has a volume of $L_{box}^{3}=(40$ kpc$)^3$. 
Since a resolution similar to that of the 2D simulation is computationally prohibitive, we run a small
series of 3D simulations at increasing resolutions, starting from the same initial conditions, and
checked numerical convergence in order to assess the robustness of our results. 

Another difference with respect to the 2D setup concerns the grid, which in this case
is not fixed. We exploit instead the reduced computational cost of the AMR technique.
Initially, the refinement criterium is geometry-based:
 maximum resolution\footnote{The maximum resolution can be computed as $L_{box} \times 0.5^{levelmax} \sim 78$ pc, where $levelmax=9$
 is the maximum refinement level used in this work.} is assigned to 
a sphere of size $\sim 8$ kpc, chosen to be larger than the whole cloud.
The resulting initial grid is nested, with the computational region located outside the maximum resolution
sphere refined at intermediate resolution; the minimum resolution, $\sim 2.5$ kpc, is adopted in the outermost regions of the box.
At later times, the refinement strategy is discontinuity based, i.e. resolution is increased in regions presenting a steep
density contrast. 
  
As initial conditions, in a reference frame whose origin is placed at the centre of the cold gas cloud,
the hot gas is moving along the x-axis at a velocity of 200 km/s. 
The cold cloud is assumed to be homogeneous and characterised by a uniform temperature, as the external, hotter medium, representing the ICM. 
The initial parameters characterising both the cloud and the ICM are the same as the ones used in the 2D simulation (Table~\ref{ParSim}). 

Also in this case, radiative cooling and heating are taken into account.
The cooling function implemented in RAMSES takes into
account both atomic (i.e., due to H and He) and metal cooling (see \citealt{few14}).
The contribution from metals at temperatures above $10^4$ K is accounted for through a 
fit of the difference between the cooling rates calculated at solar metallicity and those at zero
metallicity using the photoionisation code
CLOUDY (\citealt{fer98}). At lower temperatures, metal fine-structure cooling rates are from \cite{ros95}.
The effects of a UV background are also taken into account. 
The RAMSES cooling routine includes functional fits for the photo-heating and photoionisation rates of the \cite{haa96} ionizing
background spectrum,  as formulated in \cite{the98}.
The adopted normalization of the UV flux and the slope of the spectrum are the same as used for the 2D simulation. 
The minimum temperature is set to $T_{min}=5000 K$.
In our 3D simulations, we do not take into account thermal conduction.

Two-dimensional density and temperature maps of the 3D simulations are shown in Fig. ~\ref{3dmaps} at the same time-steps displayed in Fig.~\ref{Snapshot} for the 2D simulation, to ease a direct comparison.
Each map represents a section of the computational domain in the x-y plane.

The lack of thermal conduction makes the cloud much more prone to the effects of Kelvin-Helmholtz instabilities,
leading to a faster evolution, with respect to the 2D simulation \citep[as expected according to][]{armillotta16}, with significant morphological changes. Throughout the evolutionary time considered here, from the initially spherical shape, the cold cloud evolves into a typical umbrella-like or jellyfish-like shape,
already seen in previous studies \citep[e.g.,][]{mari11,kwak11,scanna}.

At $t=0.5$ Gyr, the x-y cross section of the cloud
has significantly thinned.  Cold gas is being continuously ablated from the cloud and mixed with the hotter ICM. 
As visible in the density and temperature maps, 
the material present at the upper and lower edges of the cloud appears stretched outwards.
The thin, extended filaments driven by the shear instabilities  cause the coolest gas to extend for nearly 20 kpc along the y-direction. As a consequence of such effects, the originally cold matter of the cloud undergoes
a substantial re-distribution in that, beside being reduced to a thinner shell,
the volume
occupied by the cold gas also increases along the z-direction. This goes along with a strong decrease of
the central density.
At this time, The wake of the cloud is still tenuous ($n \le 10^{-4}cm^{-3}$) and hot
($T\sim 10^{6.5}$ K). 

At t= 1 Gyr, the turbulent mixing of the cold and hot layers in the wake has lead to an 
increase of the filling factor of hot gas at $T\sim 10^6$ K. 
Although the distribution of the coolest gas is still preserving
a substantial degree of symmetry with
respect to the x- axis, at this epoch its stretching is extreme.
A prominent, elongated bar is visible in the front of the cloud,
whereas the average density of the wake has globally increased.
Similar protruding structures in the front of cold clouds moving fast in a very hot medium 
have been found also in previous 2d \citep{kwak11}
and 3d \citep{scanna} hydrodynamic studies, 
as well as the presence of a  few cold cloudlets detaching from  the wake.

However, Fig.~\ref{cold} shows that, notwithstanding the considerable morphological evolution, a large fraction ($\simeq$ 75\%) of the original cold gas is preserved.  
Despite the notoriously greater difficulty in obtaining numerical convergence in AMR simulations 
than in uniform grid ones (e.g., \citealt{cal15} and references therein), Fig.~\ref{cold} shows also
that the cold mass decrease is rather robust with respect to resolution. 

Hence, the results of the 3D simulations confirm that under the conditions chosen in this work,
after 1 Gyr the majority of the initial cloud mass is still cold. The 2D and 3D simulations lead to the same verdict, i.e. the considered gas cloud can survive a travel of 1~Gyr (corresponding to $\simeq$200~kpc in our simulations) through an ambient mimicking the LVC, loosing only between 5\% and 25\% of the original cold gas.
We note also that, convolving the end-state of both simulations with the beam the VLA data by A15, to simulate their observations, the resulting images are broadly compatible with the observed H{\sc i} density maps presented by these authors.

\section{Star formation in stripped clouds}
\label{starf}

Having established that a compact pressure-supported cloud similar to SECCO~1 can probably survive $>1$~Gyr long journeys within the Virgo ICMs it remains to be explored what may have triggered the onset of the current episode of star formation just a few tens of Myr ago. 

\begin{figure}
   \includegraphics[width=\columnwidth]{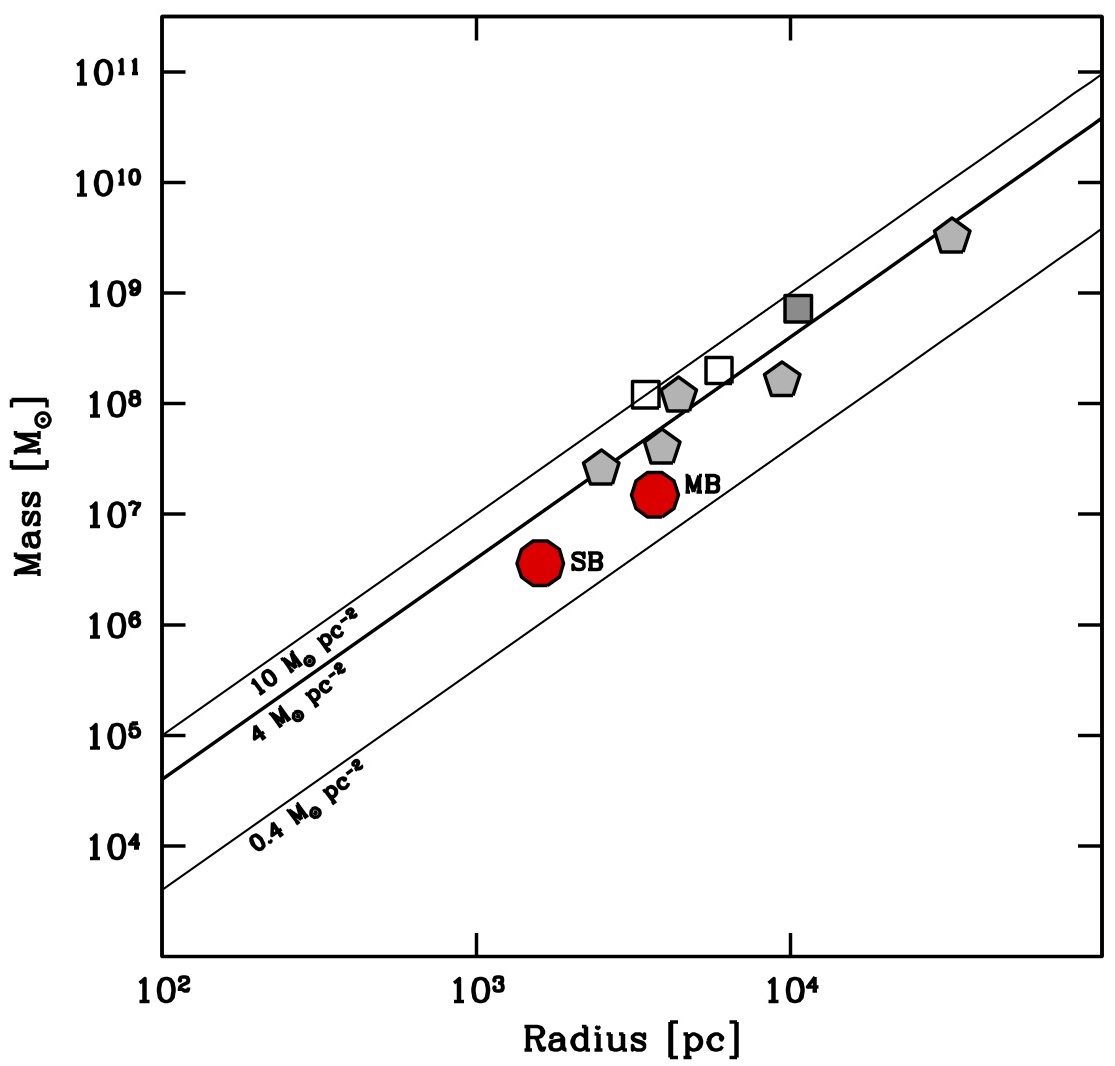}
     \caption{Size vs. mass relationship for starless or nearly starless HI clumps, following \citet{burk}. Squares are from \citet{jano15}, pentagons from \citet{can_dark}, grey filled symbols are clouds where star formation has been detected. SECCO~1 MB and SB are plotted as large red filled circles and labelled. Their H{\sc i} radii and masses used in this plot are from A15. The lines are loci of constant column density, where 4-10~$M_{\sun}~{\rm pc}^{-2}$ is the range of critical the surface density for the formation of molecular hydrogen. Gas clouds should lie above these lines to be able to ignite star formation.} 
        \label{size}
    \end{figure}

In Fig.~\ref{size} we plot the gaseous components of MB and SB in the size - column density diagram, together with other H{\sc i} clumps with or without detected star-formation inside, following \citet{burk}. These authors observe that these pressure-supported clouds are near the critical density threshold to form molecular hydrogen (4-10~M$_{\sun}$~pc$^{-2}$) and, consequently, are likely on the verge of igniting star formation (indeed, some of them are already forming stars). They suggest that pressure fluctuations and tidal shear may drive the clouds to cross the critical density line, eventually triggering a star formation event. 

We note that SECCO~1 MB and SB are not far from the other dark or almost-dark H{\sc i} clouds plotted in Fig.~\ref{size} but lie below the generally accepted critical density lines. However this may be due to the smearing effect of the VLA beam of A15 observations. The two gas clouds associated to SECCO~1 may be fragmented in smaller sub-clumps where the actual density is, in fact, above the threshold. The extreme isolation of SECCO~1 makes tidal shear unlikely to have triggered star formation in this case and it is unclear what can cause a significant pressure fluctuation in the surroundings of the system. Perhaps, during the travel, the density of the cloud may slowly increase, finally leading to the critical conditions to ignite star formation. This should be investigated with future simulations including self-gravity.

In this context, it is worth considering the relevant results of the thorough study by \citet{kap09}. These authors performed  N-body/hydrodynamical simulations, with radiative cooling, star formation and stellar feedback, of the ram pressure stripping of a large disc galaxy in a cluster environment. The final outcome of the simulations varies depending on the relative velocity of the galaxy and the ICM and on the density of the ICM, but, in general they find (a) that the gas stripped from the disc can reach distances of $\simeq 400$~kpc from the parent galaxy in just 500~Myr, (b)
that star formation occurs at any time in the compact gas cloud that forms in this extended ram pressure wake, and (c) that some of these clouds, at any distance from the parent galaxy along the wake, are still forming stars in the last 50~Myr of the 500~Myr-long simulation, and hence may host H{\sc ii} regions\footnote{Note, however, that all the star-forming knots that are active in the last 50~Myr in the \citet{kap09} simulations are associated with an underlying older population that is not observed in SECCO~1, at least within the limits of currently available HST and ground based photometry.}. These long-lived star-forming knots are broadly comparable to SECCO~1. 

However there is a remarkable difference between the results presented by \citet{kap09}
and the observations of SECCO~1. In all the simulations the star-forming knots are part of a huge and structured wake including several dense gas clumps (with or without recent star formation) and an extended cometary tail of stars formed in the wake at any epoch. At odds with this scenario, the ALFALFA survey did not detect any additional UCHVC or H{\i} structure \citep[like, e.g., the one discussed by][]{tom} in the surroundings of SECCO~1 \citep{adams}\footnote{ALFALFA covers most of the extension of the Virgo cluster in the sky and its full velocity range, and it is sensitive down to $\sim 3.0\times10^{7}~M_{\sun}$ at that distance \citep{ALFAhuang,ALFAhalle}.}. Moreover S17 searched for small star forming knots similar to SECCO~1 on deep Next Generation Virgo cluster Survey \citep[NGVS,][]{ferra} optical images, sampling a region of about $1.3\times1.3$~Mpc (in projection) around SECCO~1, finding only a few candidates that do not seem related to SECCO~1.

\subsection{Candidate dark clouds in Virgo}

It is worth noting that there are three additional UCHVCs from the \citet{adams} sample that B15a searched for stellar counterparts (finding none) that (a) are projected onto the sky area covered by the Virgo cluster, (b) have systemic velocity within the range covered by Virgo galaxies, and (c) have physical properties, as seen by ALFALFA, very similar to the original cloud where we discovered SECCO~1 (HVC~274.68+74.0). These are HVC~277.25+65.14-140 (SECCO Field~C), HVC~298.95+68.17+270 (Field~M), and HVC~290.19+70.86+204 (Field~X), see B15a and Be16. 
Adopting $D=17.0$~Mpc and using Eq.~7 by \citet{adams} we find that, if they are indeed located in the Virgo cluster, two of them (HVC~277.25+65.14-140 and HVC~290.19+70.86+204) have also H{\sc i} masses very similar to HVC~274.68+74.0 (a few $10^{7}~M_{\sun}$, in that scale), while HVC~298.95+68.17+270 would have $M_{HI}\simeq 4\times 10^{8}~M_{\sun}$.
These are good candidate to be pressure-supported clouds like SECCO~1 in which star formation has not (yet?) started \citep[see also][and references therein]{tay13,tay17,burk}. Note, however, that they  cannot be related to SECCO~1 since they are more than 1.5~Mpc away from it, in projection. It is also worth stressing that while a clustering of blue stars like SECCO~1 MB would have been easily identified by B15 or S15,  a sparser group (e.g., smaller or more dispersed than SB) could be impossible to identify in broad-band optical/UV images. Deep H${\alpha}$ relatively wide-field imaging would be the only efficient way to exclude the presence of  isolated H{\sc ii} regions or young stellar systems with $M_{\star}< 10^4~M_{\sun}$ within these clouds, whose typical projected size is $\sim 5\arcmin-10\arcmin$ \citep{adams}.

Finally we note that in both the UCHVC samples of \citet{galfa} and \citet{adams} there are a few additional  objects whose position in the sky and radial velocity is compatible with membership or association to the Virgo cluster\footnote{In particular, limiting to those lying in the range $180.0\degr \le {\rm RA}\le 195.0\degr$ and $0.0\degr \le {\rm Dec} \le 20.0\degr$, HVC~292.94+70.42+159, HVC~295.19+72.63+225, in the \citet{adams} list, and 183.0+04.4-112, , 184.8+05.7-092, and 188.9+14.5+387, in the \citet[][]{galfa} list. In the \citet{galfa} sample we considered only clouds classified as {\em Candidate Galaxies} (CG).} . Hence, it is quite possible that a small but non-negligible population of compact, dark and isolated pressure-confined H{\sc i} clouds living in or around this cluster has been already observed.

\section{Summary and Conclusions}
\label{conc}

We presented a thorough analysis and discussion of the available data on the almost-dark stellar system SECCO~1, with additional results from the MUSE observations shown in Be17a,b and an analysis of the stellar content of the system, based on coupling the stellar photometry from HST-ACS data and MUSE spectroscopic data. We also presented the results from deep H${\alpha}$ imaging of the surroundings of the system obtained with OSIRIS@GTC. By means of a set of dedicated 2D and 3D hydrodynamical simulations  we studied the evolution of a gas cloud mimicking SECCO~1 that is pressure-supported within the ICM of the Virgo substructure in which SECCO~1 is embedded. 
The main observational results a can be summarised as follows:

\begin{enumerate}

\item{} All the H${\alpha}$ emitting sources identified in SECCO~1 by Be17a are unequivocally classified as H{\sc ii} regions based on line ratio diagnostics. We do not see any trace of Supernova remnants in our MUSE data. The velocity dispersion within ionised nebulae is similar to that generally observed in typical H{\sc ii} regions.

\item{} We provide evidence that the anomalously high [OIII]/H${\beta}$ ratios observed in a few sources of SECCO~1 by B17a are not due to differences in metallicity with the other sources but to differences in the ionisation status. Hence all the oxygen abundance estimates of H{\sc ii} regions in SECCO~1 obtained in Be17a are reliable and robust.

\item{} Spectral line diagnostics suggest that the stars that are ionising the gas in the  H{\sc ii} regions should be very young (age$\la 4$~Myr).

\item{} The analysis of the CMD of SECCO~1 with the H{\sc ii} regions cross-identified implies that most of the star formation in SECCO~1 likely occurred within the last 35~Myr, with no perceivable sign of old or intermediate-age populations associated to the system. However, deeper photometry is required to completely rule out the presence of an old population, especially if it is as metal rich as the H{\sc ii} regions or more.

\item{} The inconsistency between the broad-band colours of some stars associated with H{\sc ii} regions and the temperature of the ionising sources estimated from nebular lines suggests that the position of these sources in the CMD is likely affected by the nebular spectrum, making the interpretation of the CMD less straightforward \citep[see][]{reines}. 

\item{} By means of deep H${\alpha}$ imaging of a field of projected radius $\simeq 20$~kpc we were able to identify only one robust candidate star-forming region (with a GALEX counterpart) possibly associated with SECCO~1, confirming the remarkable isolation of this system. Spectroscopic follow up is required to ascertain the actual nature of the candidate. 

\end{enumerate}

Based on the above points, on the numerical simulation presented here as well as on previous results from the literature we can make some final considerations on the nature of SECCO~1:

\begin{itemize}

\item{} While the structural and kinematical parameters of SECCO~1, taken at face value, are consistent with fundamental scaling laws of galaxies (namely the $M_V$ vs. log(r$_h$) relation and the baryonic Tully-Fisher relation, see Sect.~\ref{disc}, for discussion), its oxygen abundance is a factor of $\ga 10$ larger than the typical galaxy of the same stellar mass. This implies that the system most likely originated from a ram-pressure or tidal stripping event that removed a pre-enriched gas cloud from a galaxy with $M_{\star}\ga 10^8~M_{\sun}$.
The virial ratio of the gas cloud indicates that it is not gravitationally bound but it is confined by the external pressure.

\item{} The main physical properties of SECCO~1 are very similar to those of {\em fireballs} \citep{yoshi,fuma}. However, star formation is ignited in fireballs virtually as soon as the gas from which they form is stripped from the parent galaxy by ram-pressure, while SECCO~1 is extremely isolated and star formation is currently ongoing.

\item{} We identified another possible site of origin for the stripped gas cloud that formed SECCO~1. It is an interacting triplet of dwarf galaxies (VCC~322, VCC~319, VCC~334) that is closer to SECCO~1 in projection than previously suggested candidates, and whose members have oxygen abundance very similar to SECCO~1.

\item{} Independently of the actual site of origin, the stripped gas cloud that formed SECCO~1 should have travelled for $\sim 1$~Gyr within the ICM of LCV before the onset of the currently ongoing star formation episode
\citep[or, perhaps, forming stars more or less continuously, like some of the gas blobs in the simulations by][]{kap09}. The hydrodynamical simulations that we performed suggest that a pressure-supported gas cloud like SECCO~1 can probably survive such a long time in the considered environment, in agreement with previous results from the literature. The process that lead the cloud to cross the critical density to ignite star formation is unclear and cannot be investigated with the simulations presented here. The current unremarkable environment of SECCO~1 suggests some kind of slow internal mechanism as the most plausible driver of this density evolution.

\item{} We note that three of the \citet{adams} UCHVC included in the SECCO survey are compatible with belonging to the Virgo cluster (or to one of its substructures). They are quite similar to SECCO~1 and, assuming a distance of D$=17.0$~Mpc, also the H{\sc i} mass of two of them is virtually the same as SECCO~1. Since B15a found no sign of  stellar counterpart in their very deep images, they are good candidate pressure-supported gas cloud that have not yet ignited star formation \citep[like those considered by][see discussion and references therein]{burk}. These clouds, if they are indeed located in Virgo, are just above the sensitivity limit of ALFALFA (in H{\sc i} mass). Hence they may be the tip of the iceberg of a large population of compact gas clouds floating in galaxy clusters (see S17 for further discussion). This hypothesis is supported by the fact that, in addition to those considered here, a handful of the UCHVCs identified by \citet{galfa} and \citet{adams} also have positions and velocities compatible with association with the Virgo cluster.

\end{itemize} 

One of the most remarkable features of SECCO~1 is that it comes in two pieces. There are only two possible paths to this outcome, in the very likely hypothesis that stripping is at the origin of the overall system. MB and SB were independently stripped from the same parent galaxy and/or fragmented during the stripping event, and managed to fly together for more than 1~Gyr. On the other hand one can imagine that they were a single cloud that fragmented in two pieces, perhaps at the onset of the current star formation episode, due, e.g., to mechanical feedback from Supernovae. If fragmentation occurred 35~Myr ago, the observed projected distance would imply a velocity difference of $\ga 180\kms$.  

Indeed, the effect of SN feedback on SECCO~1 is probably the most interesting issue to be studied, e.g., by means of numerical simulations, as it will not only help to clarify the origin of the system but also to understand its fate. Would the stars remain bound together as hypothesised by \citet{yoshi} for the fireballs? Is the spatial offset between SB and the AGC~229490 cloud due to gas ejection by Supernova driven winds or to the natural lag of the stars that are immune from the friction exerted by the ICM on the gas? But if this is the case why is the same offset not seen in MB? The two pieces should have similar direction of motion in the plane of the sky, hence the lag between stars and gas should be similar. The offset between SB and AGC~229490 implies velocities in the range $\sim 50-500\kms$ for ejection times between 35~Myr and 4~Myr ago. Such velocities seem incompatible with a lag due to ram pressure, since, according to our simulation, in that timescale it would have produced a negligible velocity difference between the gas and the stars ($<1.0\kms$). Unfortunately the uncertainties on the actual status of the medium in which SECCO~1 is immersed and the limitations of our simulations prevent to draw firm conclusions from this kind of arguments. It is also interesting to note that the total baryonic mass of SECCO~1 MB is above the threshold generally considered sufficient to retain SN ejecta \citep[see, e.g.][and references therein]{ws12}. 

Finally, it can be speculated that even if SECCO~1 will be further fragmented into individual  H{\sc ii} regions (or
compact groups of them), once the ionising sources have exploded as SNe, the stars of lower mass within individual regions may remain bound as star clusters with size $\la 5$~pc and mass $\sim 10^2-10^4~M_{\sun}$. If this is the case and SECCO~1 in not an unique system, {\em intra cluster open clusters} should be found free-floating in Virgo (and in other galaxy clusters as well), in addition to globular clusters \citep{icgc}.

\section*{Acknowledgements}

We are grateful to an anonymous referee for precious comments and suggestions.
We are grateful to Blakesley Burkhart for providing the data used in her paper and for useful discussions.
G.B. gratefully acknowledges the financial support by the Spanish Ministry of Economy 
and Competitiveness under the Ramon y Cajal Program (RYC-2012-11537) and the grant AYA2014-56795-P.
F. C. acknowledges funding from the INAF PRIN-SKA 2017 program 1.05.01.88.04. We also acknowledge the {\em Accordo Quadro INAF-CINECA (2017)}, for the availability of high performance computing resources and support.

This paper uses data from SDSS - DR12. Funding for the Sloan Digital Sky Survey (SDSS) has been provided by the Alfred P. Sloan Foundation, the Participating Institutions, the National Aeronautics and Space Administration, the National Science Foundation, the U.S. Department of Energy, the Japanese Monbukagakusho, and the Max Planck Society. The SDSS Web site is http://www.sdss.org/.
The SDSS is managed by the Astrophysical Research Consortium (ARC) for the Participating Institutions. The Participating Institutions are The University of Chicago, Fermilab, the Institute for Advanced Study, the Japan Participation Group, The Johns Hopkins University, Los Alamos National Laboratory, the Max-Planck-Institute for Astronomy (MPIA), the Max-Planck-Institute for Astrophysics (MPA), New Mexico State University, University of Pittsburgh, Princeton University, the United States Naval Observatory, and the University of Washington.
The authors acknowledge the CNAF institute of the Italian Institute for Nuclear Physics (INFN)
  for providing part of the computational resources and support required for this work.
This research has made use of the SIMBAD database, operated at CDS, Strasbourg and 
of the NASA/IPAC Extragalactic Database (NED) which is operated by the Jet Propulsion Laboratory, California Institute of Technology, under contract with NASA.









\bsp	
\label{lastpage}
\end{document}